\documentclass[final,3p,times,sort&compress,doublespacing,fleqn,authoryear]{elsarticle}

\usepackage{multirow,multicol,setspace}

\usepackage{amsthm,amssymb,array,float,latexsym,amsmath,amssymb,amsbsy,eurosym,theorem,amsfonts,mathrsfs,bbm,verbatim}
\usepackage{booktabs}
\usepackage{natbib}

\usepackage{rotating}
\usepackage{tabularx}
\usepackage[table]{xcolor}
\usepackage[]{graphicx}
\usepackage{color}

\renewcommand{\b}[1]{\boldsymbol{#1}} 

\renewcommand{\c}[1]{\mathcal{#1}}

\renewcommand{\o}[1]{\overline{#1}}

\renewcommand{\t}[1]{\widetilde{#1}}

\newcommand{\p}{\partial}

\newcommand{\bnull}{\b0}

\renewcommand{\d}{\mbox{d}}

\newcommand{\Grad}{\ensuremath{\mbox{Grad}}}

\newcommand{\Div}{\mbox{Div}}

\newcommand{\dyad}{\otimes}

\newcommand{\rfr}{_{\text{\tiny{\!\! 0}}}} 
\newcommand{\crn}{_{{\tiny{t}}}} 

\newcommand{\trns}{{}^{\mathrm{\scriptsize t}}}

\newcommand{\invtrns}{{}^{\text{\footnotesize -}\text{\scriptsize t}}}

\newcommand{\sth}{\{\bullet\}}

\newcommand{\stH}{\{\circ\}}

\newcommand{\jmp}[1]{[\hspace{-2.0pt}[ {#1} ]\hspace{-2.0pt}]}
\newcommand{\avg}[1]{\{\hspace{-2.0pt}\{ {#1} \}\hspace{-2.0pt}\}}

\newcommand{\lravg}[1]{\{\!\{{#1}\}\!\}}

\newcommand\norm[1]{\left\lVert#1\right\rVert}

\makeatletter
\def\ps@pprintTitle{%
   \let\@oddhead\@empty
   \let\@evenhead\@empty
   \let\@oddfoot\@empty
   \let\@evenfoot\@oddfoot
}
\makeatother

\begin{document}

\clearpage


\clearpage

\begin{frontmatter}

\title{Generalized interfaces via weighted averages\\ for application to graded interphases at large deformations}

\author[erlangen]{S.~Saeb}
\author[erlangen]{S.~Firooz}
\author[erlangen,glasgow]{P.~Steinmann}
\author[bilkent]{A.~Javili\corref{cor}}
\ead{ajavili@bilkent.edu.tr}
\address[bilkent]{Department of Mechanical Engineering, Bilkent University, 06800 Ankara, Turkey}
\address[erlangen]{Institute of Applied Mechanics, University of Erlangen-Nuremberg, Egerland Str. 5, 91058 Erlangen, Germany}
\address[glasgow]{Glasgow Computational Engineering Centre, School of Engineering, University of Glasgow, Glasgow G12 8QQ, United Kingdom}
\cortext[cor]{Corresponding author.}

\begin{abstract}
Finite-thickness \emph{interphases} between different constituents in heterogeneous materials are often replaced by a zero-thickness \emph{interface} model.
Commonly accepted interface models intuitively assume that the interface layer is situated exactly in the middle of its associated interphase.
Furthermore, it has been reported in the literature that this assumption is necessary to guarantee the balance of angular momentum on the interface.
While the interface coincides with the mid-layer of a uniform interphase, we argue that this assumption fails to sufficiently capture the behavior of graded or inhomogeneous interphases.
This contribution extends the formulation of the general interface model to account for arbitrary interface positions.
The issue of angular momentum balance on general interfaces is critically revisited.
It is proven that the interface position does not necessarily have to coincide with the mid-layer in order to satisfy the angular momentum balance.
The analysis here leads to a unique definition of the controversially discussed interface configuration.
The presented general interface model is essentially based upon the \emph{weighted average operator} instead of the commonly accepted \emph{classical average operator}.
The framework is geometrically exact and suitable for finite deformations.
The significance of the interface position is demonstrated via a series of examples where the interface position is identified based on a full resolution interphase.
\end{abstract}

\begin{keyword}
Interphase, General interface model, Weighted average, Interface elasticity, Cohesive interface
\end{keyword}

\end{frontmatter}

\section{Introduction}\label{sec:intro}

\noindent
The study of interphases has been a longstanding subject in the context of mechanics and physics of multi-phase materials.
In particular, the advent of nano-structured materials with large area-to-volume ratio has stimulated numerous investigations on interphase effects due to its prominent contributions to material properties~\citep{Wu2004,Kari2008,Li2011a}.
It is widely recognized that interphases between the constituents of complex materials exhibit different behavior compared to their surrounding bulk.
This difference, for instance, is due to atomic lattice mismatch, poor mechanical or chemical adherence, coatings, surface contamination or debonding~\citep{Torquato1995}.
To model the behavior of such finite-thickness interphases, various zero-thickness imperfect interface models have been proposed.
The term ``imperfect'' here implies that certain fields experience a jump across the interface and they are not continuous anymore.
In this manuscript, we limit our focus to mechanical problems where the fields of interest are the displacement and traction fields.
Seminal works of~\cite{Hashin1991,Hashin1991b,Hashin1990}, \cite{Benveniste2001} and \cite{Rubin2004} have meticulously investigated the correlation between the interphase properties and the interface conditions.

There are two major approaches to determine interface properties; the phenomenological approach and the asymptotic analysis approach.
In the \emph{phenomenological approach}, the interface characteristics such as tangential and orthogonal resistance are determined separately as parts of the model.
Among the well-established interface models based on the phenomenological approach, the elastic interface model, the cohesive interface model and the general interface model are the most widely adopted ones.
The \textit{elastic interface model}, which can be viewed as an extension of the surface elasticity theory~\citep{Gurtin1975}, assumes a continuous displacement whereas it allows for a traction jump across the interface.
One of the major shortcomings of the Gurtin--Murdoch theory was that the interface was modeled as a zero thickness layer that rendered no resistance against bending.
This issue was addressed by Steigmann and Ogden~\citep{Steigmann1997,Steigmann1999} where they generalized the Gurtin--Murdoch model via incorporating the flexural resistance as well as curvature effects into the elastic interface model.
The \textit{cohesive interface model}~\citep{Dudgale1960,Barenblatt1962,Needleman1987a} preserves the traction continuity while allowing for a displacement jump across the interface.
The material behavior in the cohesive zone models is described by a traction-separation law.
The cohesive interface model has experienced a prolific growth as a research area and has been extensively utilized in multiple disciplines~\citep{ortiz1999,Gasser2003,Achenbach1989,Achenbach1990}.
Both the elastic and cohesive interface models can be interpreted as the two limit cases of a \textit{general interface model}~\citep{Hashin2002,Benveniste2006,Javili2017c} where both displacement and traction jumps across the interface are admissible.
The material behavior for the general interface models is, in general, a combination of the tangential and orthogonal behavior.
The tangential behavior is similar to the elastic interface model and the orthogonal behavior resembles the cohesive interface model.
Accordingly, the general interface model is capable of recovering both the cohesive (spring-type) and elastic (stress-type) interface models.
Table~\ref{tab:refs2} gathers major and recent contributions on interface models in the context of mechanical problems.

\begin{table}[b!]
\setstretch{1.2}
   \caption{Major contributions in interface models.}
   \label{tab:refs2}
   \small 
   \centering 
   \begin{tabular}{|p{0.02\textwidth}||p{0.9\textwidth}||}
   
   \midrule
    \multirow{5}{*}{\rotatebox{90}{elastic}} & \footnotesize{
    \citet{Gurtin1975,Gurtin1978,Dell'Isola1987,Klarbring1991,Zhong1997,Steigmann1999,Benveniste2001,Sharma2003,Sharma2004,Sun2004,Yang2004,Fried2005,Dingreville2005,Duan2005,Duan2005a,Huang2006,Benveniste2007,Chen2007a,Chen2007,Quang2007,Park2007,Park2008,Quang2008,Mogilevskaya2008,Wang2010a,Altenbach2011,Monteiro2011,Mogilevskaya2010,Kushch2011,Javili2013b,Han2018,Mogilevskaya2010a}
    }
    \\
    \midrule
    \multirow{6}{*}{\rotatebox{90}{cohesive}} & \footnotesize{
    \citet{Dudgale1960,Barenblatt1962,Bose1974,Theocaris1978,Lene1982,Benveniste1984,Benveniste1985,Needleman1987a,Takahashi1988,Karihaloo1988,Klarbring1998,Achenbach1989,Achenbach1990,Hashin1992,Qu1993,Jun1993,Zhong1997,ortiz1999,Alfano2001,Wells2001,Moes2002,Gasser2003,Hansbo2004,Mergheim2006,Charlotte2006,Fagerstrom2006,VandenBosch2008,Brassart2009,Ngo2010,Park2013,Tu2014}
    }
    \\
    \midrule
    \multirow{3}{*}{\rotatebox{90}{general}} & \footnotesize{
    \citet{Bovik1994,Hashin2002,Benveniste2006,Monchiet2010,Gu2011,Gu2014,Xu2016,Javili2017c,Chatzigeorgiou2017a,Saeb2018,Javili2018a,Heitbreder2018,Saeb2019,Firooz2019d,Firooz2019b,Firooz2019a,Saeb2019a}
    }
    \\
    \midrule
    \end{tabular}
\end{table}

In the \emph{asymptotic analysis approach}, the thickness of the interphase is assumed to be a small parameter that tends to zero and the jump conditions across the interface are determined by the continuum mechanics equations and constitutive response of the interphase, see for instance~\citep{Klarbring1998,Klarbring1991,Hashin1991,Bigoni1998,Rizzoni2017,Benveniste2001,Benveniste2006,Benveniste2006a,Benveniste2010a,Baranova2020,Baranova2020a}.
Unlike the phenomenological approach, the interfacial behavior in the asymptotic approach is captured due to the existing thickness-dependent quantities and thus, no further assumption is required to define the tangential or orthogonal behavior of the interface.
The main idea in the asymptotic analysis approach is therefore to exploit the formal asymptotic expansion that is obtained via a perturbation method or Taylor series and impose that on the displacement and traction fields of the finite-thickness interphase.
This approach naturally relates the field variables within the interphase layer by the corresponding field variables in the adjacent bulk materials.
Hence, the displacement or traction jump across the interface become a function of the displacement and traction fields in the bulk.
To represent the interphase layer precisely, a key requirement is to maintain accuracy as the interphase thickness increases.
Employing an expansion of order zero, one obtains a perfect interface model with vanishing displacement and traction jumps.
Considering higher orders in the expansion yields an imperfect interface model with a transmission condition involving the displacement and traction vectors and their derivatives, see~\cite{Lebon2010,Lebon2011,Rizzoni2013} among others.
While the majority of the contributions on asymptotic analysis have studied first-order expansions within the framework of elasticity~\citep{Hashin1991,Hashin2002,Benveniste2006,Duan2005a,Gu2008,Gu2011,Gu2014,Rizzoni2013,Serpilli2019}, a few higher order models also exist in the literature~\citep{Benveniste2006a,Benveniste2010a,Baranova2020,Rizzoni2014a}.

Although numerous contributions in the literature have studied imperfect interfaces, almost all of them have disregarded the problem of the interface position.
Unlike the general interface model, the interface position does not play a role in determining the interfacial behavior for the cohesive and the elastic interface models. 
The classical \emph{cohesive interface models} adopt the standard traction-separation law which relates the average cohesive tractions to the displacement jump.
Thus, the displacement jump determines the cohesive interface response.
The interface position plays no role in evaluating the displacement jump and consequently, it does not contribute to the interfacial behavior.
For the \emph{elastic interface model}, due to the vanishing displacement jump, the interface always coincides with its two adjacent bulk materials and thus, the interface position becomes irrelevant.
For the \emph{general interface model} though, in contrast to the cohesive and elastic interface models, the position of the interface plays a crucial role in determining the interfacial behavior.
To the best of our knowledge, this issue has been disregarded in almost all of the contributions dealing with the general interface models, see for instance~\cite{Hashin1990,Hashin1991,Benveniste2006,Gu2014,Xu2016,Javili2017c,Heitbreder2018}.
Commonly accepted general interface models intuitively assume that the interface is situated exactly in the middle of its associated interphase so as to satisfy the interface angular momentum balance.
This notion has been partially investigated in~\citep{Vossen2013,Ottosen2015,Heitbreder2017} where they demonstrate that in order to satisfy the balance of angular momentum, the traction vector must be collinear with the displacement jump.
In addition,~\cite{Mosler2011a} examined the interface position for a class of general interfaces and explicitly concluded that the only admissible position for the interface to satisfy angular momentum balance is the mid-layer, see also~\citep{Ottosen2016}.
Similarly,~\cite{Javili2017c} define the interface on the mid-layer so as to satisfy the interface balance of moments on the interface.
Further investigations on interface position in the context of weak discontinuities include~\citep{Hansbo2002,Hansbo2004,Mergheim2006}.
However, none of these works deal with the general interface model.

Almost all contributions employing the asymptotic analysis approach have assumed that the interface is located in the middle of its corresponding interphase, possibly due to the fact that these works often correspond to uniform interphases and thus, considering a mid-layer position for the interface is intuitive and justifiable.
To the best of our knowledge, there exist only two contributions where asymptotic analysis is carried out on a graded interphase.
\cite{Benveniste2007a} conducted a three-dimensional asymptotic analysis to model a curved graded interphase by an interface model in the context of heat conduction.
In their analysis, they assumed that the interface is located at the mid-layer of the interphase and, indeed, they observed and reported that their asymptotic analysis rendered a somewhat inconsistent behavior for a highly graded interphase and their results were not in good agreement with exact solutions when the interphase was not very thin.
Later, \cite{Lebon2011} carried out an asymptotic analysis on graded interphases but provided no benchmark examples or numerical study to verify their proposed model.
Apart from neglecting the complexity of the interphase structure, the asymptotic analysis approach is usually corresponds to small-strain linear elasticity.
At large deformations, however, the position of the interface plays a crucial role that is investigated in the current contribution.

\begin{figure}[b!]
  \centering
  \includegraphics[width=1.0\textwidth]{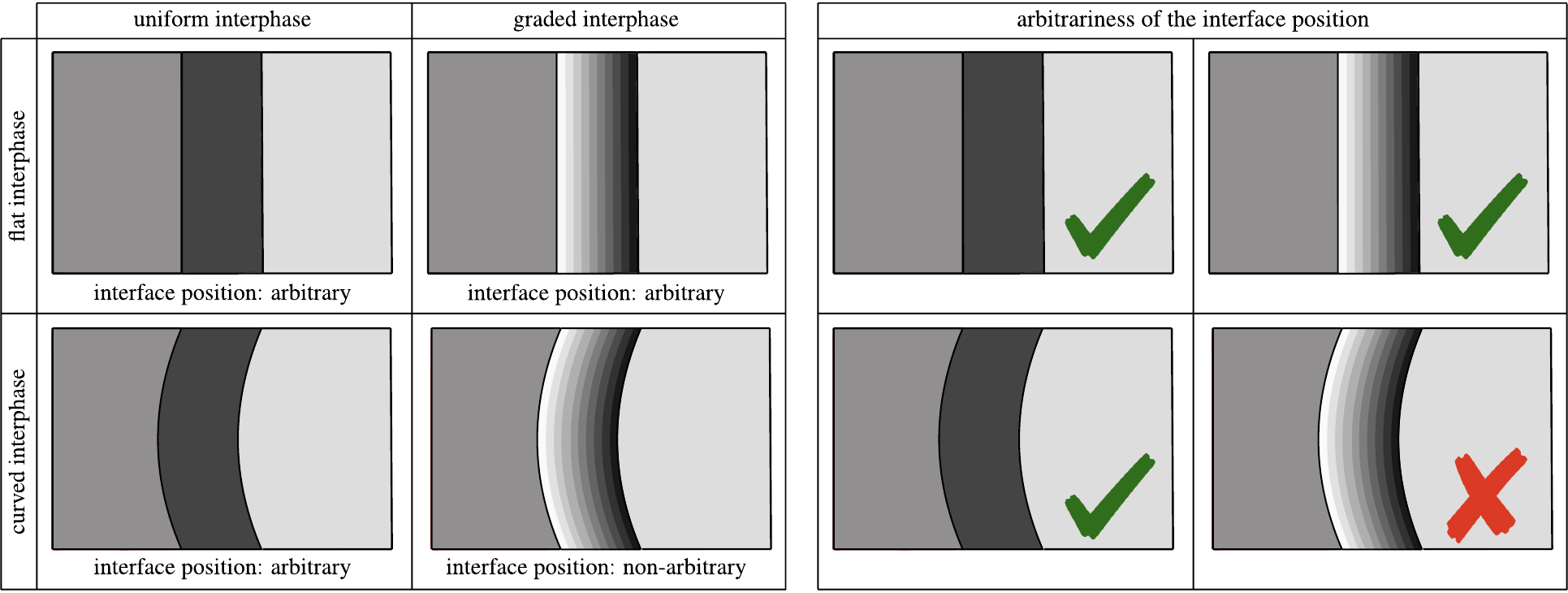}
  \caption{
  Schematics illustration of the role of the interface position in capturing the interphase behavior.
  For a flat interphase or a uniform graded interphase, the interface position is arbitrary and may be assumed to coincide with the mid-layer for convenience.
  For a curved graded interphase, the interface position is no longer arbitrary and plays a significant role in recovering the interphase behavior.
  This observation is supported by numerical examples in Section~\ref{sec:discussion} and the four cases are gathered in Fig.~\ref{fig:compare}.
  }
  \label{fig:motivation}
\end{figure}

\begin{figure}[b!]
  \centering
  \includegraphics[width=0.95\textwidth]{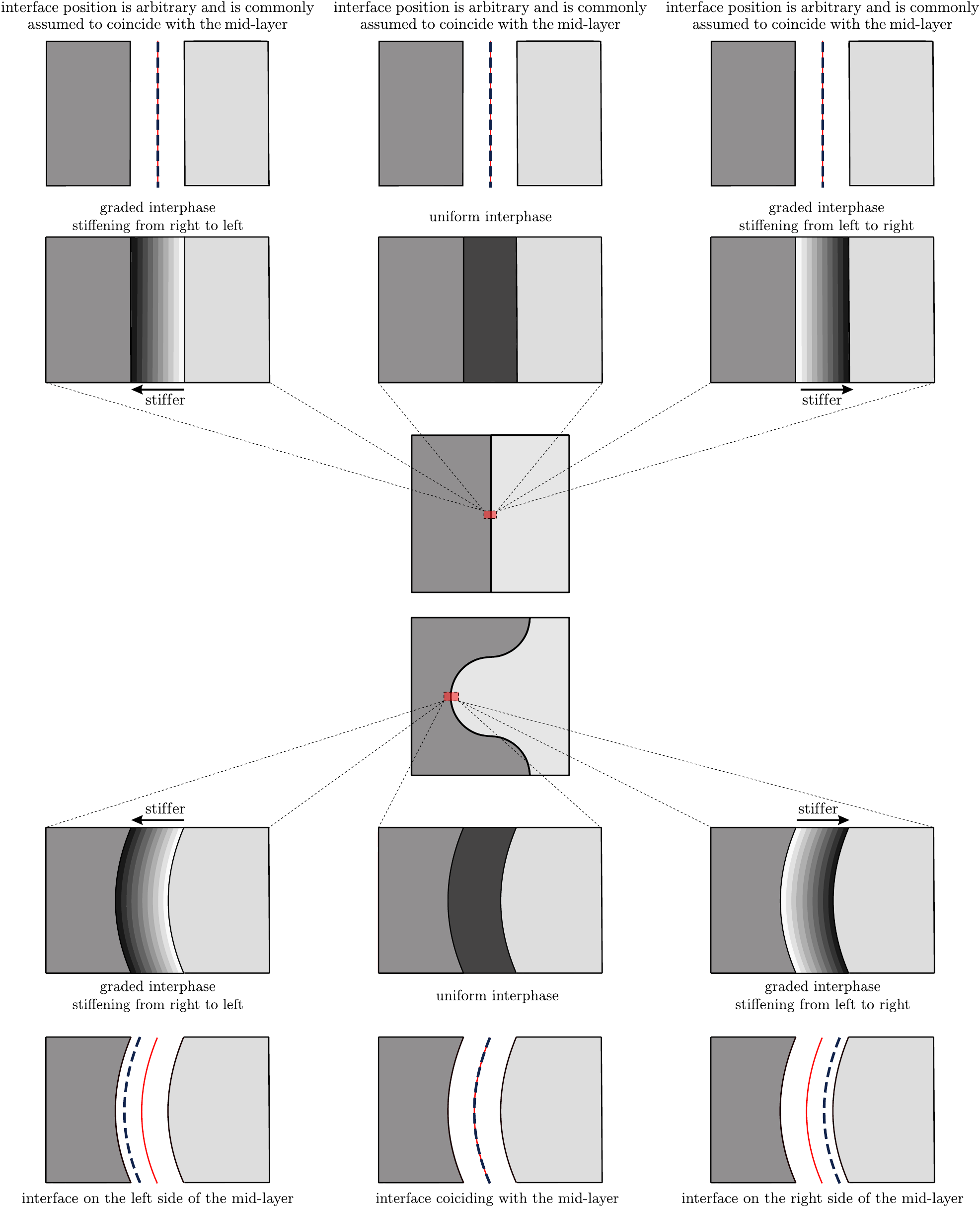}
  \caption{
  Schematics illustration of a proper interface position for various interphase micro-structures and geometries.
  To properly capture a graded interphase behavior, the interface position should be chosen on the stiffer side of the interphase but not on the mid-layer.
  For any interphase, a unique set of parameters for the general interface model exists that minimizes the error of the interface model compared to its associated full resolution interphase.
  }
  \label{fig:intro}
\end{figure}

In this contribution, we carry out a comprehensive study on the role of the interface position in capturing the interphase behavior for different interphase structures at finite deformations where geometrical and material non-linearities are present.
We demonstrate that the assumption that a general interface model coincides with the mid-layer is only justifiable for uniform interphases or for interphases with simple geometries.
Figure~\ref{fig:motivation} schematically illustrates the role of the interface position in capturing the interphase behavior for four interphases with different geometries and structures.
These micro-structures under extension are studied in Section~\ref{sec:discussion} and gathered in Fig.~\ref{fig:compare}.
For a flat interphase, a general interface with an arbitrary position can sufficiently capture the interphase effects since the interface tangential resistance plays no significant role in this case, which can be explained by the generalized Young--Laplace equation.
For a curved interphase, both tangential and orthogonal interface properties contribute to the material's response.
Nonetheless, for a homogeneous interphase, the influence of the interface position is negligible.
For a curved graded interphase, the unknown interface position is required to be calculated via minimizing the error of the interface model compared to its associated full resolution interphase.
This observation forms the basis of the current manuscript and the numerical examples shed light on our findings.
Note that the presence of graded interphases between the constituent of composite materials is very common and have been reported in many experimental studies such as~\cite{Theocaris1985a} for E-glass/fiber epoxy composites, \cite{Jayaraman1993} for E-glass/IMHS epoxy, carbon/IMHS epoxy and Kevlar-49/IMHS epoxy composites, \cite{Lutz1997} for mortar with sand inclusions and \cite{Low1994,Low1995} for graphite/fiber epoxy and carbon/fiber epoxy composites.

The assumption of general interfaces coinciding with the mid-layer is no longer valid necessarily for complex interphase structures.
Instead, a more realistic approach suggests interface positions other than the mid-layer, see Fig.~\ref{fig:intro} for a schematic illustration.
For curved interphases with graded structures, we show that the interface position cannot be constrained to the mid-layer but it must lie within the stiffer part of the interphase to capture the interphase behavior more accurately.
To do so, we establish a novel formulation for the general interface model via adopting a weighted average that allows for arbitrary choices of the interface position.
We rigorously demonstrate that the assumption of restricting the interface position to the mid-layer is merely a conjecture and any arbitrary position for the interface is admissible without violating the angular momentum balance.
In summary, the key features and contributions of this manuscript are:
\begin{itemize}
\item To propose a novel class of general interface models via introducing the weighted average and complimentary weighted average operators,
\item To present a generalized form of the interface balance of moments allowing for arbitrary interface positions,
\item To establish an extended traction-separation law accounting for the interface position,
\item To demonstrate the significance of the interface position on accurately capturing the response of complex inhomogeneous interphases.
\end{itemize}


\section{Theory}\label{sec:GII}
\noindent The purpose of this section is to formulate the theory of general interfaces accounting for weighted averages.
The differences between the given formulation and the classical one (based on the assumption that an interface quantity is the average of its two sides) is highlighted in particular.
\begin{figure}[b!]\label{Fig1}
	\centering
	\includegraphics[angle=-90,width=1\textwidth]{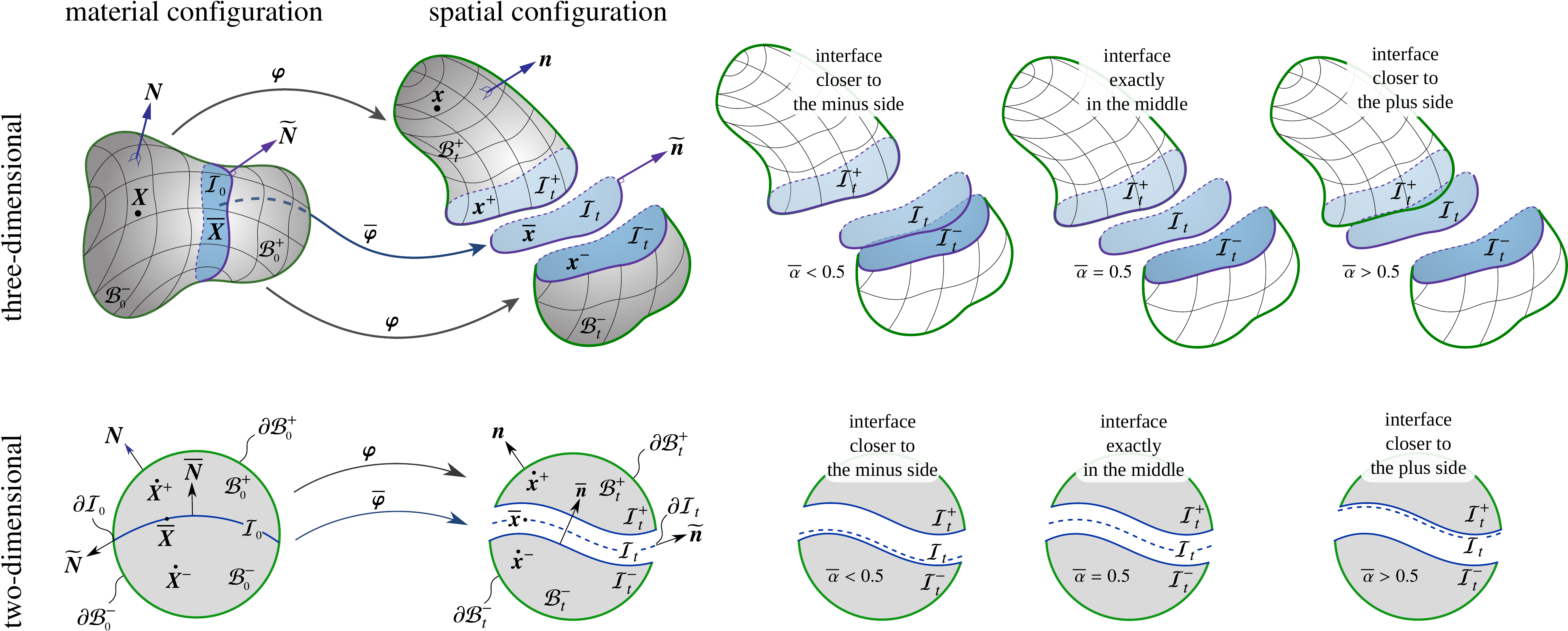}
	\caption{Deformation of a continuum body possessing an interface under finite deformations in three-dimensional (top) and two-dimensional (bottom) settings.
	The two sides of the body in the material configuration intersect at the interface $\c{I}\rfr$ with $\jmp{\b{X}}=\bnull$, dividing the bulk into two sides $\c{B}\rfr^{-}$ and $\c{B}\rfr^{+}$.
	Through the deformation $\b{\varphi}$ the displacement jump $\jmp{\b{x}}\neq\bnull$ across the interface occurs.
	The positions of the points on the interface in the spatial configuration are evaluated as a weighted average of the positions of the points on the two sides of the interface as $\o{\b{x}} = \o{\alpha} \b{x}^{+} + [1-\o{\alpha}] \b{x}^{-}$.
	Thus, for $\o{\alpha} < 0.5$, the interface stays closer to the minus side while for $\o{\alpha} > 0.5$, the interface gets closer to the plus side.
	Clearly, $\o{\alpha} = 0.5$ renders the standard assumption and the interface coincides exactly with the mid-layer.
	}
	\label{fig:interface-motion-2D}
\end{figure}
\subsection{Kinematics}
\noindent Let $\c{B}\rfr$ define a continuum body in the material configuration which consists of two disjoint subdomains, $\c{B}^{+}\rfr$ and $\c{B}^{-}\rfr$, intersecting at the interface $\c{I}\rfr$, see Fig.~\ref{Fig1}.
The boundaries of $\c{B}^{+}\rfr$ and $\c{B}^{-}\rfr$ are denoted $\p\c{B}^{+}\rfr$ and $\p\c{B}^{-}\rfr$, respectively.
The interface $\c{I}\rfr$ is a two-sided surface whose intersection with the boundaries $\p\c{B}\rfr^{-}$ and $\p\c{B}\rfr^{+}$ renders $\c{I}\rfr^{-}$ and $\c{I}\rfr^{+}$, respectively.
The three surfaces $\c{I}\rfr$, $\c{I}\rfr^{-}$ and $\c{I}\rfr^{+}$ coincide in the material configuration.
The boundary of the interface $\c{I}\rfr$ is denoted ${\p\c{I}\rfr}$.
The bulk subdomains in the spatial configuration are denoted $\c{B}^{+}\crn$ and $\c{B}^{-}\crn$ whose boundaries are $\p\c{B}^{+}\crn$ and $\p\c{B}^{-}\crn$, respectively.
The interface and its boundary in the spatial configuration are denoted $\c{I}\crn$ and ${\p\c{I}\crn}$.
Analogously, the counterparts of $\c{I}\rfr^{-}$ and $\c{I}\rfr^{+}$ in the spatial configuration are $\c{I}\crn^{-}$ and $\c{I}\crn^{+}$.
The three surfaces $\c{I}\crn^{-}$, $\c{I}\crn$ and $\c{I}\crn^{+}$ may or may not coincide in the spatial configuration.
The outward unit normals to ${\p\c{B}\rfr}$ and ${\p\c{B}\crn}$ are denoted $\b{N}$ and $\b{n}$, respectively.
The outward unit normals to ${\p\c{I}\rfr}$ and ${\p\c{I}\crn}$ are $\t{\b{N}}$ and $\t{\b{n}}$, respectively and are tangent to $\c{I}\rfr$ and $\c{I}\crn$.
The unit normal to the interface ${\c{I}\rfr}$ is denoted $\o{\b{N}}$ and the unit normal to the interface ${\c{I}\crn}$ is denoted $\o{\b{n}}$ both pointing from the minus to the plus side of the interface.

Let $\b{X}$ define the placement of a point in $\c{B}\rfr$ mapped to its counterpart in the spatial configuration $\b{x}$ in $\c{B}\crn$ via the non-linear deformation map $\b{\varphi}$ as $\b{x} = \b{\varphi}(\b{X})$.
The deformation gradient is then defined as $\b{F} := \Grad \b{\varphi}$.
The interface sides $\c{I}^{+}\crn$ and $\c{I}\crn^{-}$ are assumed to follow the bulk motions $\b{\varphi}^{+}$ and $\b{\varphi}^{-}$ and therefore, they distance from each other due to the deformation jump across the interface $\jmp{\b{\varphi}}$.
The operator $\jmp{\sth}$ represents the jump of an arbitrary quantity across the interface defined as $\jmp{\sth} := \sth^{+}-\sth^{-}$, with $\sth^{+}$ and $\sth^{-}$ being the quantity $\sth$ on the plus and minus sides of the interface, respectively.
A key parameter in establishing the general interface kinematics is the definition of the interface position in the spatial configuration.
From a physical point of view, the points on the interface may not penetrate the boundaries $\c{I}^{-}\crn$ and $\c{I}\crn^{+}$.
In other words, the interface in the spatial configuration is strictly bounded by its two sides kinematically.
The positions of the points on the interface are generally defined as the average between the positions of the points on the interface sides $\c{I}\crn^{-}$ and $\c{I}\crn^{+}$ as $\o{\b{x}}:=\avg{\b{x}}$ in which $\avg{\sth}$ is the average operator defined as $\avg{\sth} = \tfrac{1}{2}[\sth^{+}+\sth^{-}]$.
A more inclusive choice is however to allow the interface to lie on an arbitrary plane between the two sides and define the position of the points on the interface as a \textit{weighted average} of its two sides as
\begin{equation}\label{weighted-avg}
\o{\b{x}} := \avg{\b{x}}_{\o{\alpha}} \quad \text{with} \quad\avg{{\sth}}_{\o{\alpha}} = \o{\alpha} \sth^{+} + [1-\o{\alpha}] \, \sth^{-}\quad \text{,} \quad \o{\alpha} \in [0,1] \, ,
\end{equation}
with $\o{\alpha}$ being the weighting coefficient.
Thus, when $\o{\alpha} < 0.5$, the interface is closer to the minus side while for $\o{\alpha} > 0.5$, the interface lies closer to the plus side.
Clearly, for $\o{\alpha} = 0.5$, the classical definition is recovered and the interface coincides with the mid-layer.
Based on this definition, the placements of the points in the material configuration are mapped to their counterparts in the spatial configuration via $\o{\b{x}} = \o{\b{\varphi}}(\o{\b{X}})$ with $\o{\b{\varphi}}:=\avg{\b{\varphi}}_{\o{\alpha}}$.
Additionally, the interface deformation gradient is defined as $\o{\b{F}} := \o{\Grad} \o{\b{\varphi}} = \o{\Grad} \avg{\b{\varphi}}_{\o{\alpha}}= {\Grad} \avg{\b{\varphi}}_{\o{\alpha}} \cdot \o{\b{I}}$ with $\o\Grad\,\sth$ representing the interface gradient operator and $\o{\b{I}} = \b{I} - \o{\b{N}} \dyad \o{\b{N}}$.
Therefore, the interface deformation gradient is a superficial tensor possessing the property $\o{\b{F}}\cdot\o{\b{N}}=\b{0}$.

\begin{table}[h!]
\setstretch{1.8}
   \caption{Summary of some of the notations and definitions in the bulk and on the interface. The third order permutation tensor is denoted $\b{\varepsilon}$. The interface curvature is $\o{C}$.}
   \label{TAB1}
   \centering 
   \small
   \begin{tabular}{|p{0.17\textwidth}||c|c|}
   \midrule
   & bulk & interface \\
   \midrule
   \midrule
    divergence operator & $\Div{\sth} = \Grad{\sth} : \b{I}$  & $\o\Div \, {\o\sth}  = \o\Grad \, \o\sth : \o{\b{I}}$ \\\midrule
	traction            & $\b{t}\rfr$ [N/m$^{2}$]                         & $\o{\b{t}}\rfr$ [N/m]  \\\midrule
	force density         & $\b{b}\rfr$  [N/m$^{3}$]                        & $\o{\b{b}}\rfr$ [N/m$^{2}$] \\\midrule
	Piola stress        & $\b{P}$     [N/m$^{2}$]                         & $\o{\b{P}}$    [N/m]  \\\midrule
	Cauchy theorem         & $\b{t}\rfr = \b{P} \cdot \b{N}$     & $\o{\b{t}}\rfr = \o{\b{P}} \cdot \t{\b{N}}$ \\\midrule
	divergence theorem  & $\int_{\p\c{B}\rfr} \sth \cdot \b{N} \, \d A = \int_{\c{B}\rfr} \Div {\sth} \, \d V + \int_{\c{I}\rfr} \jmp{\sth} \, \d A$  & $\int_{\p\c{I}\rfr} \o{\sth} \cdot \t{\b{N}} \, \d A = \int_{\c{I}\rfr} \o{\Div}\, \o{\sth} \, \d A +  \int_{\c{I}\rfr} \o{C} \, \o{\sth}\cdot\o{\b{N}} \, \d A $  \\\midrule
	useful identity     & \multicolumn{2}{c|}{$\Div (\b a \times \b B) = \b a \times \Div \b B + \b\varepsilon : [\Grad \b a  \cdot {\b{B}}\trns]$ \quad,\quad $\b{a}$ and $\b{B}$: arbitrary first and second order tensors}  \\
    \midrule
    \end{tabular}\,
\end{table}

Next, we elaborate on the derivation of the governing equations, for an arbitrary (otherwise constant) value of $\o{\alpha}$.
The governing equations of the bulk remain indifferent to $\o{\alpha}$.
The derivation of the balance equations for the bulk is standard and is briefly discussed so as to set the stage for deriving the interface equations.
Henceforth, the problem is treated as quasi-static, for the sake of simplicity.
Table~\ref{TAB1} summarizes the notations and identities used here.

\subsection{Governing equations in the bulk}
\noindent The governing mechanical equations are the balances of forces and moments.
In order to derive the strong form of the governing equations, the static equilibrium conditions for an arbitrary cutout volume in the bulk is first established.
Consider an arbitrary cutout volume of the bulk denoted $\c{V}\rfr$ surrounded by its boundary $\p\c{V}\rfr$.
The cutout volume is chosen such that it contains no portion of the interface. 
Writing the balance of forces then furnishes
\begin{equation}\label{eq:1}
\int_{\p\c{V}\rfr} \b{t}\rfr \, \d A + \int_{\c{V}\rfr} \b{b}\rfr \, \d V = \int_{\p\c{V}\rfr} \b{P} \cdot \b{N} \, \d A + \int_{\c{V}\rfr} \b{b}\rfr \, \d V = \int_{\c{V}\rfr} \Div \b{P} \, \d V + \int_{\c{V}\rfr} \b{b}\rfr \, \d V = \b{0} \, , \quad \forall \c{V}\rfr \, .
\end{equation}
Therefore, the strong form of the balance of forces in the bulk reads
\begin{equation}\label{eq:2}
\boxed{\Div \b{P} + \b{b}\rfr = \b{0} \, .}
\end{equation}
\noindent The balance of forces is often solved numerically using the finite element method.
To do so, the strong form is recast into a weak form presented in section \ref{weak-mech}.
In addition to the balance of forces, the balance of moments should also be established.
Writing the balance of moments acting on the body with respect to an arbitrary point in space yields
\begin{equation}\label{eq:3}
\begin{aligned}
&\int_{\p\c{V}\rfr} \b{x} \times \b{t}\rfr \, \d A + \int_{\c{V}\rfr} \b{x} \times \b{b}\rfr \, \d V = \b{0}  \quad \forall \c{V}\rfr \,, 
\end{aligned}
\end{equation}
that after a few mathematical steps given in \ref{App1-1} results in the balance of moments
\begin{equation}\label{eq:4}
\boxed{\b{\varepsilon} : [\b{F} \cdot \b{P}\trns] = \b{0} \, .}
\end{equation}
\noindent In contrast to the balance of forces which is solved numerically, the balance of moments is not generally solved and does not enter the system of equations.
It is, however, preserved through choosing an objective constitutive model for the material.

\subsection{Governing equations on the interface}
\noindent Next, consider a cutout volume $\c{V}\rfr$ that contains a portion of the interface and is partitioned into two disjoint subdomains $\c{V}^{-}\rfr$ and $\c{V}^{+}\rfr$ through the interface $\c{I}\rfr$.
The boundaries of the subdomains and interface are denoted  $\p\c{V}^{-}\rfr$ and $\p\c{V}^{+}\rfr$ and $\p\c{I}\rfr$, respectively.
Similar to the bulk, the equilibrium of forces on the interface reads
\begin{equation}\label{eq:4.5}
\int_{\p\c{V}\rfr^{+}} \b{t}\rfr \, \d A + \int_{\p\c{V}\rfr^{-}} \b{t}\rfr \, \d A +  \int_{\p\c{I}\rfr} \o{\b{P}} \cdot \t{\b{N}} \, \d L + \int_{\c{I}\rfr} \o{\b{b}}\rfr \, \d A = \b{0}   \quad \forall \c{V}\rfr \,.
\end{equation}
After mathematical manipulations given in \ref{App1-2}, the balance of forces on the interface reads
\begin{equation}\label{eq:6}
\boxed{\o{\Div} \, \o{\b{P}} + \llbracket \b{P} \rrbracket \cdot \o{\b{N}} + \o{\b{b}}\rfr = \b{0} \, .}
\end{equation}
\noindent In addition to the balance of forces, the balance of moments should also be ensured.
The balance of moments acting on the interface with respect to an arbitrary point in space reads
\begin{equation}\label{eq:7}
\begin{aligned}
&\int_{\p\c{V}\rfr^{+}} \b{x} \times \b{t}\rfr \, \d A + \int_{\p\c{V}\rfr^{-}} \b{x} \times \b{t}\rfr \, \d A +  \int_{\p\c{I}\rfr} \o{\b{x}} \times \o{\b{P}} \cdot \t{\b{N}} \, \d L + \int_{\c{I}\rfr} \o{\b{x}} \times \o{\b{b}}\rfr \, \d A = \b{0}   \quad \forall \c{V}\rfr \,.
\end{aligned}
\end{equation}
Consequently, the strong form of the interface balance of moments is obtained as
\begin{equation}\label{eq:8}
\boxed{\llbracket\b{x} \times \b{P} \rrbracket \cdot \o{\b{N}} + \b{\varepsilon} : [\o{\b{F}} \cdot \o{\b{P}}\trns] - \o{\b{x}} \times \llbracket \b{P} \rrbracket \cdot \o{\b{N}} = \b{0} \, .}
\end{equation}
See \ref{App1-3} for the proof. 
The interface moments balance~\eqref{eq:8} can be further simplified.
However, as it will be demonstrated, the definition of the interface position plays a crucial role in determining the final form of the equation.
In the following, we first present the standard methodology for simplifying Eq.~\eqref{eq:8} and highlight the limitations that it will induce.
Next, we establish a novel formulation of the problem based on weighted averages that lifts the previous limitations.
Our approach covers a wider range of solutions and reduces to the standard one only as a limit.

\paragraph{Standard methodology based on common average operator}
Let $\sth$ and $\stH$ be two arbitrary interface quantities and a multiplication operator is denoted as $\star$.
The standard methodology to simplify Eq.~\eqref{eq:8} is to use the well-known identity
\begin{equation}\label{eq:9}
\begin{aligned}
	\jmp{ \sth \star \stH } = \avg{ \sth } \star \jmp { \stH } + \jmp{ \sth } \star \avg { \stH } \,.
\end{aligned}
\end{equation}
Upon using this identity, Eq.~\eqref{eq:8} is further simplified as
\begin{equation}\label{eq:9.5}
\begin{aligned}
&\llbracket\b{x} \times \b{P} \rrbracket \cdot \o{\b{N}} + \b{\varepsilon} : [\o{\b{F}} \cdot \o{\b{P}}\trns] - \o{\b{x}} \times \llbracket \b{P} \rrbracket \cdot \o{\b{N}}
\\&\quad=\lravg{\b{x}} \times \llbracket \b{P} \rrbracket\cdot\o{\b{N}} + \llbracket \b{x} \rrbracket \times \lravg{\b{P}}\cdot\o{\b{N}} + \b{\varepsilon} : [\o{\b{F}}\cdot\o{\b{P}}\trns] - \o{\b{x}} \times \llbracket \b{P} \rrbracket\cdot\o{\b{N}}
\\&\quad\quad= \llbracket \b{x} \rrbracket \times \lravg{\b{P}}\cdot\o{\b{N}} + \b{\varepsilon} : [\o{\b{F}}\cdot\o{\b{P}}\trns] + [\lravg{\b{x}} - \o{\b{x}}] \times \llbracket \b{P} \rrbracket\cdot\o{\b{N}}= \b{0} \, .
\end{aligned}
\end{equation}
The interface balance of moments is then preserved through choosing suitable constitutive laws satisfying material frame indifference.
A sufficient set of conditions is to set the three terms to zero separately as
\begin{equation}\label{conditions}
\begin{aligned}
& \llbracket \b{x} \rrbracket \times \lravg{\b{P}}\cdot\o{\b{N}} = \b{0} \quad\quad &&\text{if} \quad\quad \llbracket \b{x} \rrbracket \, \parallel \, \lravg{\b{P}}\cdot\o{\b{N}}\, , \\
& \b{\varepsilon} : [\o{\b{F}} \cdot \o{\b{P}}\trns] = \b{0}  \quad\quad &&\text{if} \quad\quad \o{\b{F}} \cdot \o{\b{P}}\trns = \o{\b{P}} \cdot \o{\b{F}}\trns \, , \\
& [\lravg{\b{x}} - \o{\b{x}}] \times \llbracket \b{P} \rrbracket\cdot\o{\b{N}} = \b{0} \quad\quad &&\text{if} \quad\quad \lravg{\b{x}} = \o{\b{x}}\, .
\end{aligned}
\end{equation}
From the condition~(\ref{conditions})$_3$, it is obvious that this approach does not allow for arbitrary choices for the interface position.
That is, any choice for the interface position other than the mid-layer violates the balance of moments.
This has been the primary reason to date for not investigating the role of the interface position for general interfaces.
We address this issue and introduce a novel formulation which paves the way for the analysis of the interface position while satisfying the balance of moments in what follows.

\paragraph{Novel formulation of the problem based on weighted averages}
The point of departure here is to define the interface position according to Eq.~\eqref{weighted-avg} as a weighted average of its two sides $\o{\b{x}} := \avg{\b{x}}_{\o{\alpha}}$, illustrated in Fig.~\ref{Fig1}.
Clearly, this definition does not, in general, satisfy the last condition in Eq.~\eqref{conditions} since $\avg{\b{x}} \ne \avg{\b{x}}_{\o{\alpha}}$.
Thus, we need to go one step further back and introduce an extended version of the identity \eqref{eq:9} to begin with.
For that, we firstly introduce the \textit{complementary weighted average}.
That is
\begin{equation}\label{weighted-avg-comp}
\begin{aligned}
&\lravg{\sth}_{\o{\alpha}}  :=  \o{\alpha} \, {\sth}^+  +[1-\o{\alpha}] \, {\sth}^- \quad&&:\quad \text{weighted average} \, , \\
&\lravg{\sth}_{1-\o{\alpha}}  := [1-\o{\alpha}] \, {\sth}^+ + \o{\alpha} \, {\sth}^- \, \quad&&:\quad \text{complementary weighted average}.
\end{aligned}
\end{equation}
\noindent Note the difference between the weighted average and the complementary weighted average operators.
Together they allow to distinguish between the quantities that are defined via different averaging rules that are otherwise impossible to identify.\footnote{This observation is somewhat reminiscent of the discussion on co-variant and contra-variant quantities that can be readily distinguished if they are expressed in a convected curvilinear coordinates system, but they are otherwise virtually impossible to identify in a classical Cartesian description.}
Next, we present a generalized version of the identity \eqref{eq:9} as
\begin{equation}\label{magic-identity-ex}
\llbracket \sth \star \{\circ\} \rrbracket= \lravg{\sth}_{\o{\alpha}} \star \llbracket \{\circ\}\rrbracket + \llbracket \sth \rrbracket \star \lravg{\{\circ\}}_{1-\o{\alpha}} \, .
\end{equation}
Evidently, the identity \eqref{magic-identity-ex} covers a wider spectrum than the widely-recognized identity \eqref{eq:9} but it also reduces to \eqref{eq:9} when $\o{\alpha}=\tfrac{1}{2}$.
Using the identity \eqref{magic-identity-ex}, the interface moment balance~(\ref{eq:8}) reads
\begin{equation}\label{eq_new}
\begin{aligned}
&\llbracket\b{x} \times \b{P} \rrbracket \cdot \o{\b{N}} + \b{\varepsilon} : [\o{\b{F}} \cdot \o{\b{P}}\trns] - \o{\b{x}} \times \llbracket \b{P} \rrbracket \cdot \o{\b{N}}
\\&\quad=\lravg{\b{x}}_{\o{\alpha}} \times \llbracket \b{P}\rrbracket \cdot \o{\b{N}} + \llbracket \b{x} \rrbracket \times \lravg{\b{P}}_{1-\o{\alpha}} \cdot \o{\b{N}} + \b{\varepsilon} : [\o{\b{F}}\cdot\o{\b{P}}\trns] - \o{\b{x}} \times \llbracket{\b{P}}\rrbracket \cdot \o{\b{N}} = \b{0} \, .
\end{aligned}
\end{equation}
Upon defining the cohesive traction as $\o{\b{t}} := \lravg{\b{P}}_{1-\o{\alpha}} \cdot \o{\b{N}}$ and inserting the interface position $\o{\b{x}} := \avg{\b{x}}_{\o{\alpha}}$ into Eq.~\eqref{eq_new}, the interface balance of moments is obtained as
\begin{equation}
\boxed{\llbracket \b{x} \rrbracket \times \o{\b{t}} + \b{\varepsilon} : [\o{\b{F}}\cdot\o{\b{P}}\trns] = \b{0} \, ,}
\end{equation}
\noindent which can be sufficiently fulfilled if each term vanishes separately as 
\begin{equation}\label{last-last-angular}
\begin{aligned}
& \llbracket \b{x} \rrbracket \times \o{\b{t}} = \b{0} \quad\quad &&\text{if} \quad\quad \llbracket \b{x} \rrbracket  \, \parallel \, \o{\b{t}}\, , \\
& \b{\varepsilon} : [\o{\b{F}}\cdot\o{\b{P}}\trns] = \b{0}  \quad\quad &&\text{if} \quad\quad \o{\b{F}} \cdot \o{\b{P}}\trns = \o{\b{P}} \cdot \o{\b{F}}\trns \, .
\end{aligned}
\end{equation}
It shall be emphasized that \textit{while the weighted average operator dictates the interface kinematics, the interface kinetics shall be defined by the complementary weighted average operator}.
That is, for instance, in the limit of $\o{\alpha} = 0$, the interface position coincides with the minus side, but, the cohesive traction consists of only the contribution from the plus side.
On the other hand, in the limit of $\o{\alpha} = 1$, the interface coincides with the plus side geometrically but only the minus side contributes to the cohesive traction.
Obviously, only for $\o{\alpha}=\tfrac{1}{2}$, the complementary weighted average coincides with the standard average and the commonly accepted form of the interface balance of moments is recovered.
Equation~\eqref{last-last-angular} is the generic form of the balance of moments on the interface, which accommodates for arbitrary values for $\o{\alpha}$, hence arbitrary interface positions between its two sides.

\begin{figure}[h!]
  \centering
    {\includegraphics[angle=-90,width=0.9\textwidth]{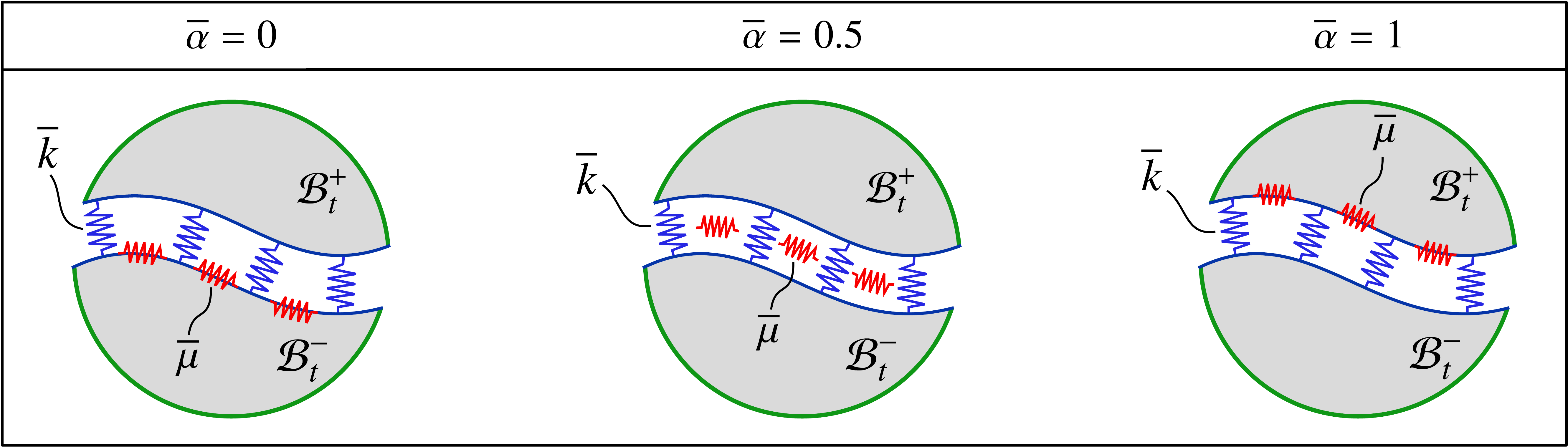}}
    \caption{Two-dimensional schematic illustration of the interface configuration for different values of $\o{\alpha}$. The blue springs represent the orthogonal resistance of the interface against opening and the red springs represent the tangential resistance against in-plane stretch.}
    \label{fig:interface-param}
\end{figure}

Following the previous works on general interfaces \citep{Javili2018a} and motivated by the structure of Eq.~\eqref{last-last-angular}, we decompose the interface response to two parts accounting for the behavior along and across the interface as
\begin{equation}\label{piola-stress-traction-separation}
\begin{split}
& \o{\b{t}} = \o k \, \llbracket \b{x} \rrbracket \, , \\
&\o{\b{P}} = \o{\mu} \, [ \o{\b{F}} - \o{\b{F}}\invtrns] + \o{\lambda} \log \o{J}\, \o{\b{F}}\invtrns \, ,
\end{split}
\end{equation}
where $\o\mu$, $\o\lambda$ and $\o k$ represent the interface material parameters governing the interface behavior.
The parameters $\o{\mu}$ and $\o{\lambda}$ represent the interface resistance against in-plane stretches and $\o{k}$ is the interface orthogonal resistance against opening.
Figure~\ref{fig:interface-param} schematically represents the interface material parameters for different interface configurations.
The interface constitutive laws~\eqref{piola-stress-traction-separation} guarantee the interface balance of moments \eqref{last-last-angular}.
The traction-separation law $\eqref{piola-stress-traction-separation}_{1}$ ensures the \emph{collinearity of the cohesive traction and the displacement jump across the interface} \citep{Vossen2013a}.
The constitutive formulations $\eqref{piola-stress-traction-separation}$ are mainly chosen for simplicity and to better understand the numerical examples and discussions in Section~\ref{sec:discussion}.

\section{Numerical illustrations and discussion}\label{sec:discussion}

\noindent This section provides numerical examples to demonstrate the significance of the interface position for accurately capturing the behavior of complex interphases.
Here we compare the general \emph{interface} model with the full resolution \emph{interphase} and identify the interface position.
Various interphase configurations are analyzed and the corresponding general interface parameters recovering the interphase behavior are obtained.
The numerical examples are devised such that they clearly highlight the importance of interface position without introducing too much complexity.
The results presented in this section are obtained using the finite element method implementation based of the weak form given in~\ref{weak-mech}.
The numerical results are limited to two-dimensional studies.
This simplification has been made for the sake of clarity and to better visualize the influence of the interface position without introducing additional complexities associated with three-dimensional problems.

\begin{figure}[h!]
  \centering
    {\includegraphics[width=1.0\textwidth]{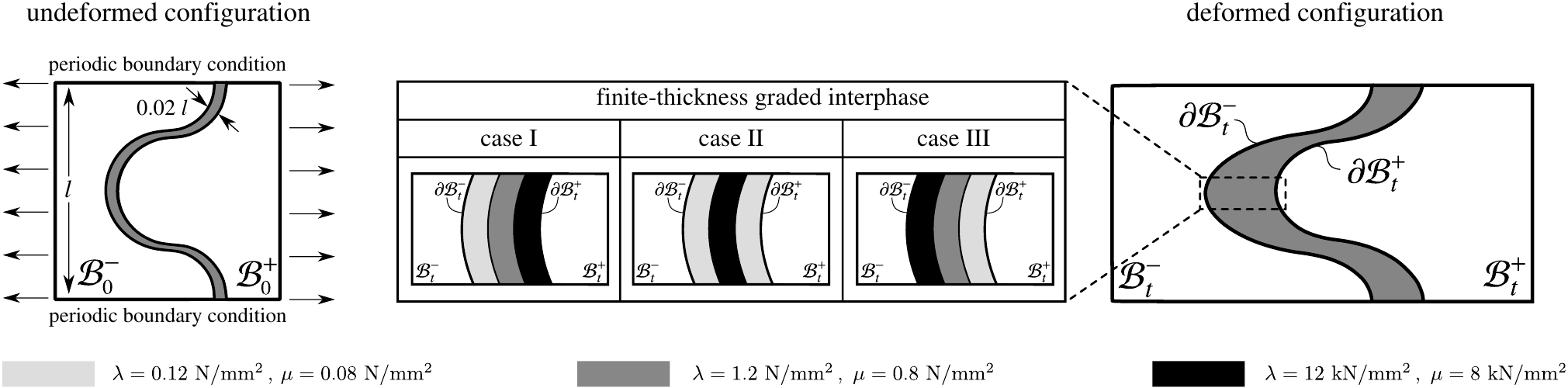}}
    \caption{
    A graded interphase between the two homogeneous subdomains $\c{B}\rfr^{+}$ and $\c{B}\rfr^{-}$, respectively.
    The thickness of the interphase is $2\%$ of the sample length.
    Three different structures are considered for the graded interphase.
    For case I, the interphase layers are assumed to become stiffer non-linearly from left to right.
    For case II, the left and right layers are assumed to be identical and more compliant than the middle layer.
    For case III, the interphase layers are assumed to become more compliant non-linearly from left to right.
    The material parameters corresponding to each layer of the interphase are given at the bottom.
    }
    \label{fig:interphase}
\end{figure}

Consider the specimen in Fig.~\ref{fig:interphase}.
The body of the specimen is partitioned by a curved \emph{finite-thickness interphase} into two homogeneous bulk subdomains denoted as $\c{B}\rfr^{+}$ and $\c{B}\rfr^{-}$ with boundaries $\p\c{B}\rfr^{+}$ and $\p\c{B}\rfr^{-}$, respectively.
The curvature of the interphase is constant and the curve is composed of four quarters of a circle.
The thickness of the interphase in the undeformed configuration is $1\%$ of the specimen side length.
An extension with a magnitude of $100\%$ is prescribed in the horizontal direction on the left and right edges of the specimen while periodic boundary conditions are imposed on the top and bottom of the specimen.
The bulk and the interphase materials behave according to the neo-Hookean constitutive law
\begin{equation}\label{bulk-material}
\b{P} = \mu \, [ \b{F} - \b{F}\invtrns] + \lambda \log J\, \b{F}\invtrns \,,
\end{equation}
with $\mu$ and $\lambda$ being the Lam{\'e }parameters.
For the bulk, $\mu\!=\!8\,\text{N}/\text{{m}}^{2}$ and $\lambda\!=\!12\, \text{N}/\text{m}^{2}$.
We consider the interphase to be composed of three different layers, each possessing its own material parameters.
The interphase composition varies, resulting in three different graded structures given in Fig.~\ref{fig:interphase} that will be examined shortly.
In the first case, the interphase gets stiffer from left to right.
In the second case, the left and right layers are identical and more compliant to the middle layer.
In the third case, the interphase layers are more compliant from left to right.
The material parameters corresponding to each layer of the interphase are provided.
The different layers of the interphase are distinguished by different shades of gray.
Light gray indicates an interphase layer that is 100 times more compliant than the bulk with the parameters $\mu\!=\!0.08\,\text{N}/\text{m}^{2}$ and $\lambda\!=\!0.12\, \text{N}/\text{m}^{2}$.
Medium gray indicates an interphase layer that is 10 times more compliant than the bulk.
Dark gray (black) interphase layer is 1000 times stiffer than the bulk.
Perfect bonding between the interphase layers and between the interphase and bulk materials is assumed.

\begin{figure}[b!]
  \centering
    {\includegraphics[width=1.0\textwidth]{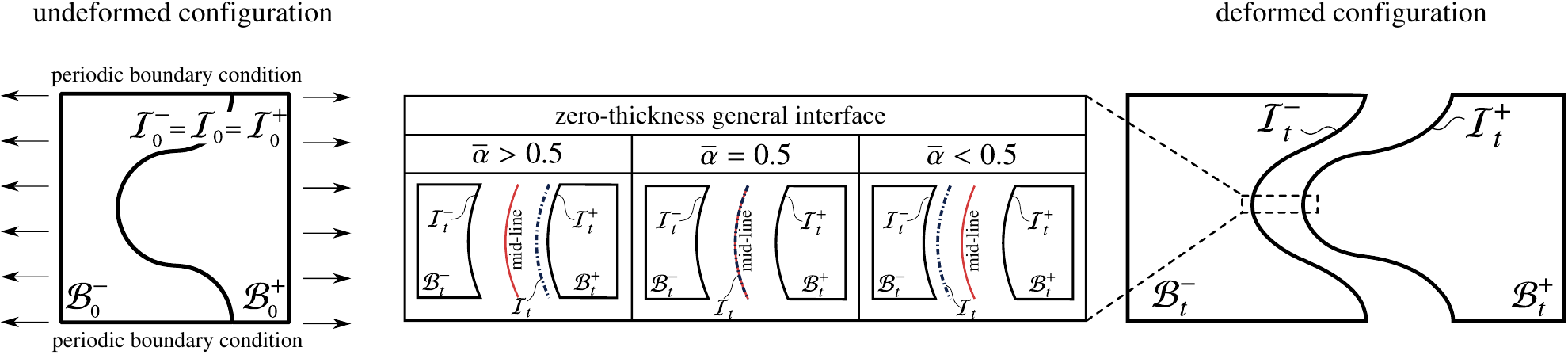}}
    \caption{
    A zero-thickness general interface model between the two homogeneous subdomains $\c{B}\rfr^{+}$ and $\c{B}\rfr^{-}$, respectively.
    Three options exist for the interface positions with respect to the mid-layer.
    The interface could be either on the left side of the mid-layer $\o{\alpha}<0.5$, coinciding with the mid-layer $\o{\alpha}=0.5$ or on the right side of the mid-layer $\o{\alpha}>0.5$.
    }
    \label{fig:interface}
\end{figure}

Next, the finite-thickness interphase is modeled via a \emph{zero-thickness general interface} $\c{I}\rfr$ with its own energetic structure.
Consider the specimen in Fig.~\ref{fig:interface} where the interface partitions the specimen into $\c{B}\rfr^{+}$ and $\c{B}\rfr^{-}$.
The curvature of the interface is constant and the curve is composed of four quarters of a circle of radius $l/4$ with $l$ being the length of the specimen.
Due to opening, in the deformed configuration, the interface sides distant from each other.
The position of each point on the interface in the deformed configuration is determined through the corresponding placements of its two sides and $\o{\alpha}$ according to $\o{\b{x}} = \avg{\b{x}}_{\o{\alpha}} = \o{\alpha} \, \b{x}^{+} + [1-\o{\alpha}] \, \b{x}^{-}$ where $\b{x}^{+}$ and $\b{x}^{-}$ correspond to the positions of the points on $\c{I}\crn^{+}$ and $\c{I}\crn^{-}$, respectively.
An extension of $100\%$ is prescribed similar to the previous case.
The interface behaves according to the constitutive law~\eqref{piola-stress-traction-separation}.
Three different strategies are possible to locate the interface.
The first and the most commonly used strategy is to assume that the interface coincides with the mid-layer or $\o{\alpha}=0.5$.
The other two strategies assume that interface is closer to one of its sides.
If the interface is closer to its minus or plus sides, then $\o{\alpha}<0.5$ or $\o{\alpha}>0.5$, respectively.
These options have not been studied in the literature, to the best of our knowledge, and are of particular interest in this contribution.
Here, we choose three cases of $\o{\alpha}=0.2$, $\o{\alpha}=0.5$ and $\o{\alpha}=0.8$.

The main objective of this numerical study is to find the interface parameters $\o\mu$, $\o{k}$, $\o{\alpha}$ that accurately recover the behavior of a graded interphase such that a one-to-one correspondence between the three interphase structures in Fig.~\ref{fig:interphase} and the three interface models in Fig.~\ref{fig:interface} is obtained.
To do so, we define an error-like parameter in order to evaluate the agreement of the results obtained by the interphase and general interface models as 
\begin{equation}
\c{E} (\o\mu, \o{k}, \o{\alpha}) = \norm{\b{x}\rvert_{\c{I}\crn^{+}} - \b{x}\rvert_{\p\c{B}\crn^{+}}} + \norm{\b{x}\rvert_{\c{I}\crn^{-}} - \b{x}\rvert_{\p\c{B}\crn^{-}}} \ ,
\end{equation}
which measures the difference between the motion of the points on the two sides of the \emph{interphase} (i.e. ${\p\c{B}\crn^{-}}$ and ${\p\c{B}\crn^{+}}$) and the motion of the points at the two sides of the \emph{interface} (i.e. ${\c{I}\crn^{-}}$ and ${\c{I}\crn^{+}}$) when the interphase is replaced by a general interface model.
Thus, smaller values of $\c{E}$ indicate a better agreement between the results obtained by the interphase and the general interface model.
For each case of interphase composition in Fig.~\ref{fig:interphase}, an extensive parametric study is carried out in order to find the interface parameters $\o\mu$, $\o{k}$, $\o{\alpha}$ such that $\c{E}$ is minimized.

\begin{figure}[h!]
  \centering
    {\includegraphics[width=1.0\textwidth]{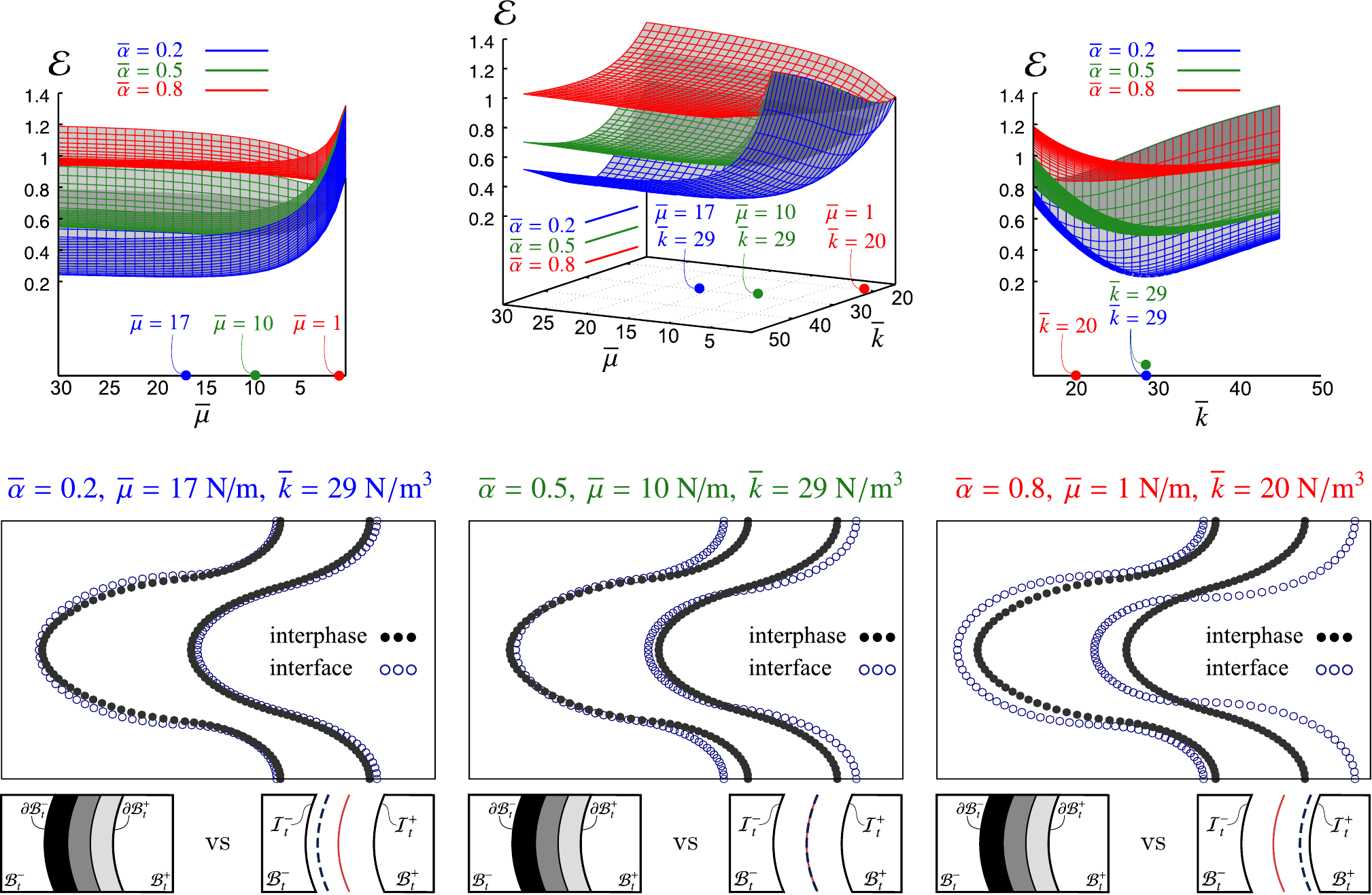}}
    \caption{
    Illustration of the influence of various interface parameters in capturing the interphase behavior corresponding to case I in Fig~\ref{fig:interphase}.
    The top plots show the norm of $\c{E}$ for a broad range of interface parameters $\o{k}$ and $\o{\mu}$ and for three different interface positions $\o{\alpha}=0.2$, $\o{\alpha}=0.5$ and $\o{\alpha}=0.8$.
    The bottom figures compare the deformations obtained by the interphase (case I) against the deformations obtained by the general interface model with different interface positions.
    }
    \label{fig:example1}
\end{figure}

\begin{figure}[h!]
  \centering
    {\includegraphics[width=1.0\textwidth]{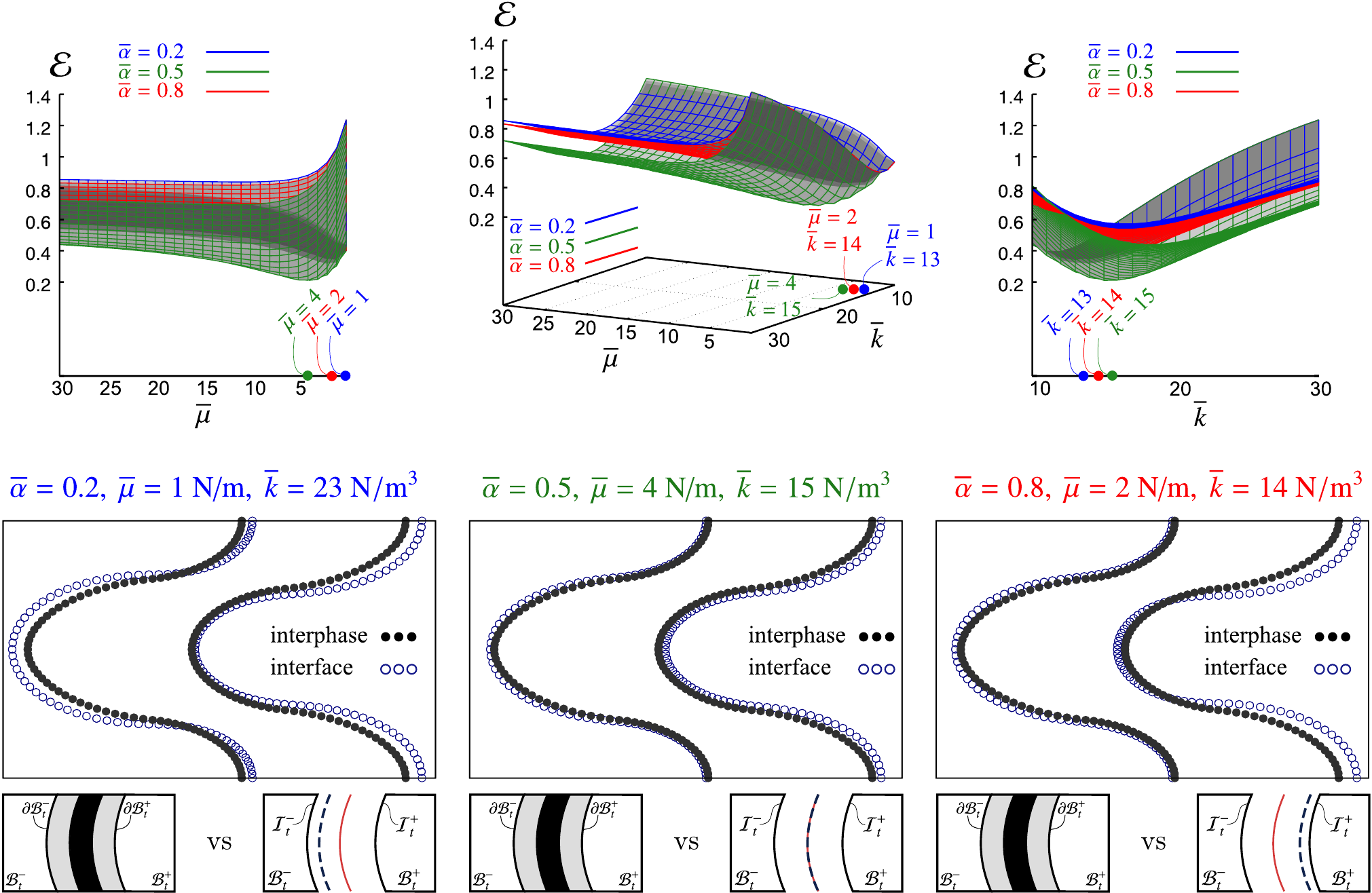}}
    \caption
    {
    Illustration of the influence of various interface parameters in capturing the interphase behavior corresponding to case II in Fig~\ref{fig:interphase}.
    The top plots show the norm of $\c{E}$ for a broad range of interface parameters $\o{k}$ and $\o{\mu}$ and for three different interface positions $\o{\alpha}=0.2$, $\o{\alpha}=0.5$ and $\o{\alpha}=0.8$.
    The bottom figures compare the deformations obtained by the interphase (case II) against the deformations obtained by the general interface model with different interface positions.
    }
    \label{fig:example2}
\end{figure}

\begin{figure}[h!]
  \centering
    {\includegraphics[width=1.0\textwidth]{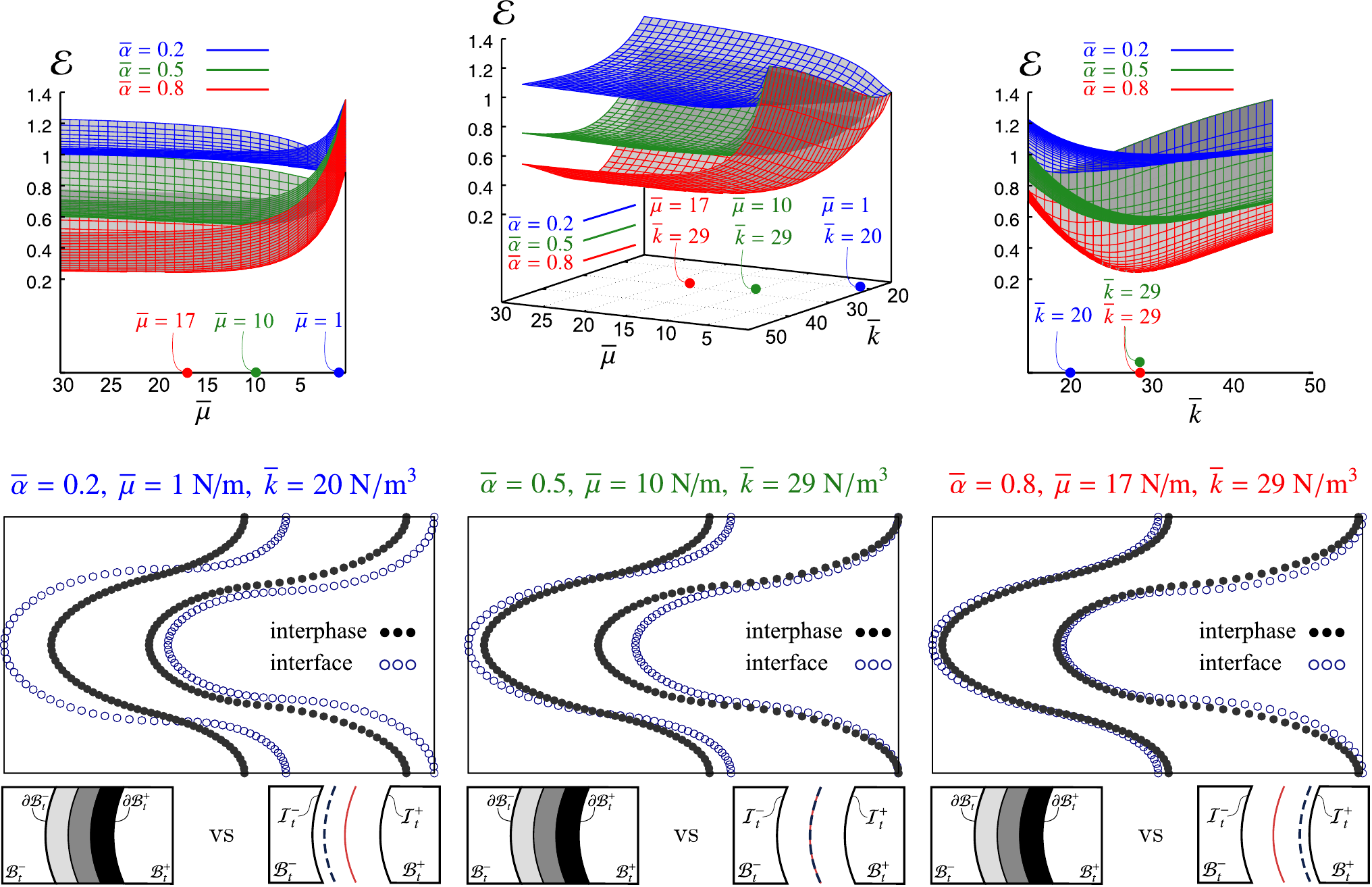}}
    \caption{
    Illustration of the influence of various interface parameters in capturing the interphase behavior corresponding to case III in Fig~\ref{fig:interphase}.
    The top plots show the norm of $\c{E}$ for a broad range of interface parameters $\o{k}$ and $\o{\mu}$ and for three different interface positions $\o{\alpha}=0.2$, $\o{\alpha}=0.5$ and $\o{\alpha}=0.8$.
    The bottom figures compare the deformations obtained by the interphase (case III) against the deformations obtained by the general interface model with different interface positions.
    }
    \label{fig:example3}
\end{figure}

Figures~\ref{fig:example1}--\ref{fig:example3} illustrate how well the general interface model captures the behavior of the three graded interphase structures shown in Fig.~\ref{fig:interphase}.
The graphs in each figure show the norm of $\c{E}$ for a broad range of interface parameters $\o{k}$ and $\o{\mu}$ and for three different interface positions $\o{\alpha}=0.2$, $\o{\alpha}=0.5$ and $\o{\alpha}=0.8$.
For better illustration, different views are provided.
The blue, green and red surfaces represent the results associated with $\o{\alpha}=0.2$, $\o{\alpha}=0.5$ and $\o{\alpha}=0.8$, respectively.
The interface parameters $\o{k}$ and $\o{\mu}$ leading to the minimum value of $\c{E}$ are highlighted as dots on each graph.
The figures at the bottom compare the deformations of the domain with the interphase or the general interface model.
The comparisons are given for three interface positions $\o{\alpha}=0.2, 0.5, 0.8$ and the optimal interface parameters $\o{k}$, $\o{\mu}$.
That is, the points on ${\p\c{B}\crn^{+}}$ and ${\p\c{B}\crn^{-}}$ are compared against the points on ${\c{I}\crn^{+}}$ and ${\c{I}\crn^{-}}$ in the deformed configuration.
The solid points show the results corresponding to the interphase layer and the hollow points render the results corresponding to the interface model.

Figure~\ref{fig:example1} exhibits the results corresponding to the case I in Fig.~\ref{fig:interphase}.
For this case, the graded interphase is assumed to become stiffer from left to right with the left layer parameters $\mu\!=\!0.08\,\text{N}/\text{m}^{2}$ and $\lambda\!=\!0.12\, \text{N}/\text{m}^{2}$, the middle layer parameters $\mu\!=\!0.8\,\text{N}/\text{m}^{2}$ and $\lambda\!=\!1.2\, \text{N}/\text{m}^{2}$ and the right layer parameters $\mu\!=\!8\,\text{kN}/\text{m}^{2}$ and $\lambda\!=\!12\, \text{kN}/\text{m}^{2}$.
Among the three interface positions associated with $\o{\alpha}=0.2, 0.5, 0.8$, we observe that $\o{\alpha}=0.2$ in combination with the interface material parameters $\o{\mu}=\!17\,\text{N}/\text{m}$ and $\o{k}=\!29\,\text{N}/\text{m}^{3}$ results in the least error and closest response to the interphase.
This example implies that \emph{to properly capture a graded interphase behavior, the interface position should be chosen on the stiffer part of the interphase and not on the mid-layer}.
Figure~\ref{fig:example2} shows the results corresponding to the case II in Fig.~\ref{fig:interphase}.
For this case, the left and right side of the graded interphase are assumed to be identical possessing the parameters $\mu\!=\!0.08\,\text{N}/\text{m}^{2}$ and $\lambda\!=\!0.12\, \text{N}/\text{m}^{2}$ and the mid-layer is stiffer than the other two with the parameters $\mu\!=\!8\,\text{kN}/\text{m}^{2}$ and $\lambda\!=\!12\, \text{kN}/\text{m}^{2}$.
Among the three interface positions, $\o{\alpha}=0.5$ in conjunction with the interface parameters $\o{\mu}=\!4\,\text{N}/\text{m}$ and $\o{k}=\!15\,\text{N}/\text{m}^{3}$ leads to the least error.
This observation seems somewhat intuitive due to the symmetry of the interphase structure.
Therefore, not every graded interphase requires interface positions other than the mid-layer.
More precisely, \emph{the interface position to accurately capture a symmetrically graded interphase coincides with the mid-layer}.
An immediate consequence of this observation is that \emph{for a uniform interphase, the interface coincided with the mid-layer}.
Figure~\ref{fig:example3} illustrated the results corresponding to the case III in Fig.~\ref{fig:interphase}.
For this case, the graded interphase is assumed to become more compliant from left to right with the left layer parameters $\mu\!=\!8\,\text{kN}/\text{m}^{2}$ and $\lambda\!=\!12\, \text{kN}/\text{m}^{2}$, the middle layer parameters $\mu\!=\!0.8\,\text{N}/\text{m}^{2}$ and $\lambda\!=\!1.2\, \text{N}/\text{m}^{2}$ and the right layer parameters $\mu\!=\!0.08\,\text{N}/\text{m}^{2}$ and $\lambda\!=\!0.12\, \text{N}/\text{m}^{2}$.
Among the three interface positions, the numerical results reveal that $\o{\alpha}=0.8$ in combination with the interface parameters $\o{\mu}=\!17\,\text{N}/\text{m}$ and $\o{k}=\!29\,\text{N}/\text{m}^{3}$ yields the least error.
This observation further supports the conclusion that in order to properly capture a graded interphase behavior, the interface position should be chosen on the stiffer portion of the interphase.

\begin{figure}[b!]
  \centering
    {\includegraphics[width=1.0\textwidth]{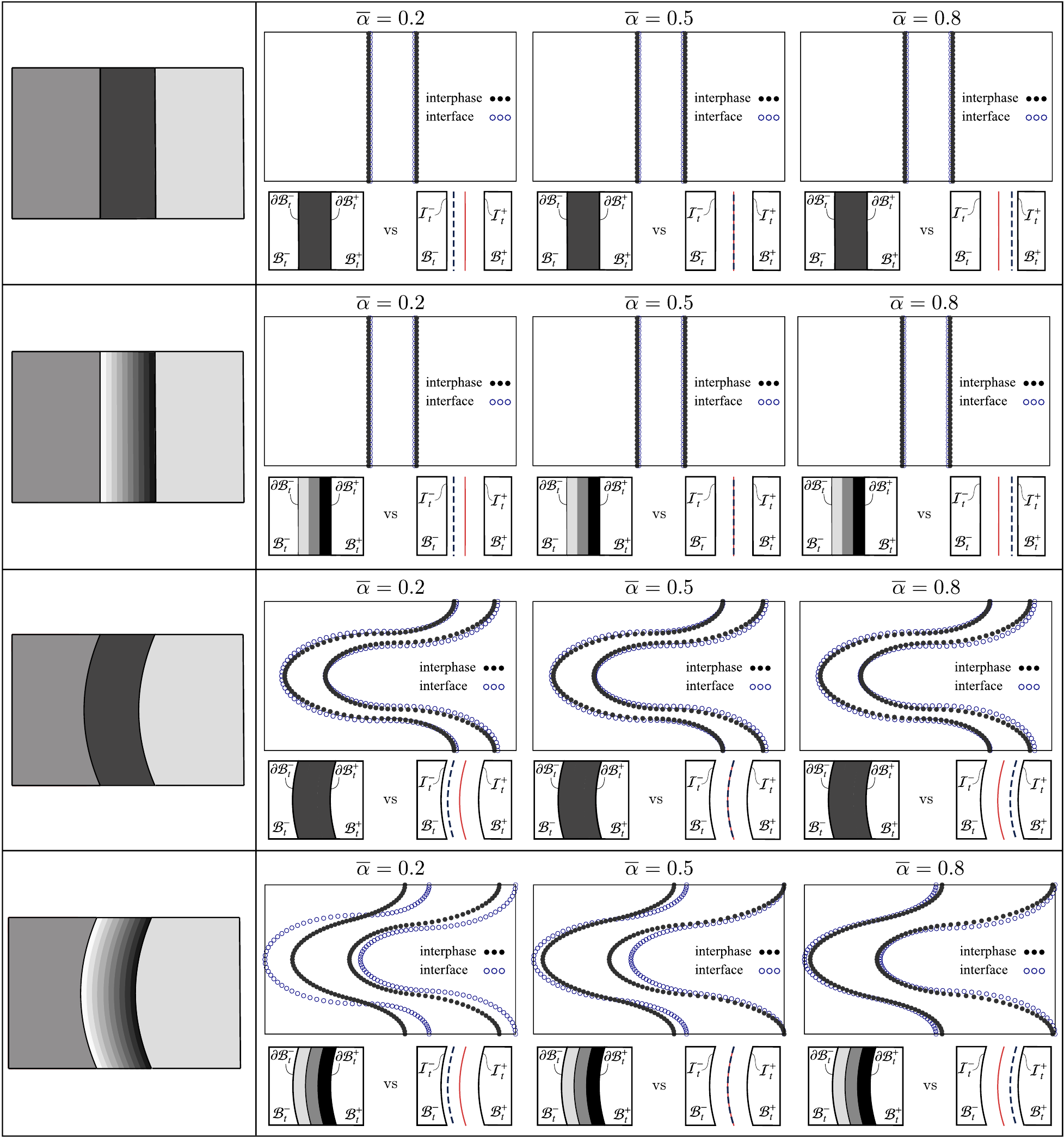}}
    \caption{
    Comparison between different interphase structures and proper general interface model to capture the interphase behavior for the four cases introduces in Fig.~\ref{fig:motivation}.
    The points illustrate the deformation of the sides of the bulk materials in the adjacency to the interface/interphase.
    The solid points correspond to the deformation obtained by the interphase and the hollow points correspond to the deformation obtained by the interface.
    The bottom figures compare the deformations obtained by the interphase (case I) against the deformations obtained by the general interface model with different interface positions.
    }
    \label{fig:compare}
\end{figure}

It is important to emphasize that for all the numerical results, the set of interface parameters $\o{\alpha}$, $\o{\mu}$, $\o{k}$ leading to the smallest $\c{E}$ is unique.
Furthermore, for each $\o{\alpha}$ an extensive study for a broad range of $\o{\mu}$ and $\o{k}$ is carried out.
Obviously, in order to obtain the most accurate set, a broader range for $\o{\alpha}$ should have been taken into account, that we have omitted for the sake of brevity.
Another conclusion that can be drawn from the examples is that neither the cohesive nor the elastic interface model alone can properly recover the response of an interphase.
Overall, this numerical investigation reveals the immense versatility of the general interface model and at the same time sheds light on its potential to model the overall behavior of multi-phase materials.

Figure~\ref{fig:compare} compares the four different cases illustrated in Fig.~\ref{fig:motivation} and clearly depicts the role of the interface position on the material response.
The results render the deformation of identical adjacent bulk materials in the presence of a finite-thickness interphase in between, similar to Figs.~\ref{fig:example1}--\ref{fig:example3}.
For the interphases with flat geometry and/or uniform micro-structure, an excellent agreement is observed between the results obtained from the full resolution interphase and the general interface model, regardless of the interface position.
This observation implies that for these cases,  the interface position is arbitrary.
On the other hand, for a curved graded interphase different interface positions lead to different results.
It is observed that interface positions corresponding to $\o{\alpha}=0.2$ and $\o{\alpha}=0.5$ cannot sufficiently capture the interphase behavior whereas $\o{\alpha}=0.8$ predicts the interphase behavior accurately.
Note that the graded interphase structure for this case, similar to Fig.~\ref{fig:example3}, stiffens from left to right.

\begin{figure}[t!]
  \centering
    \includegraphics[width=1.0\textwidth]{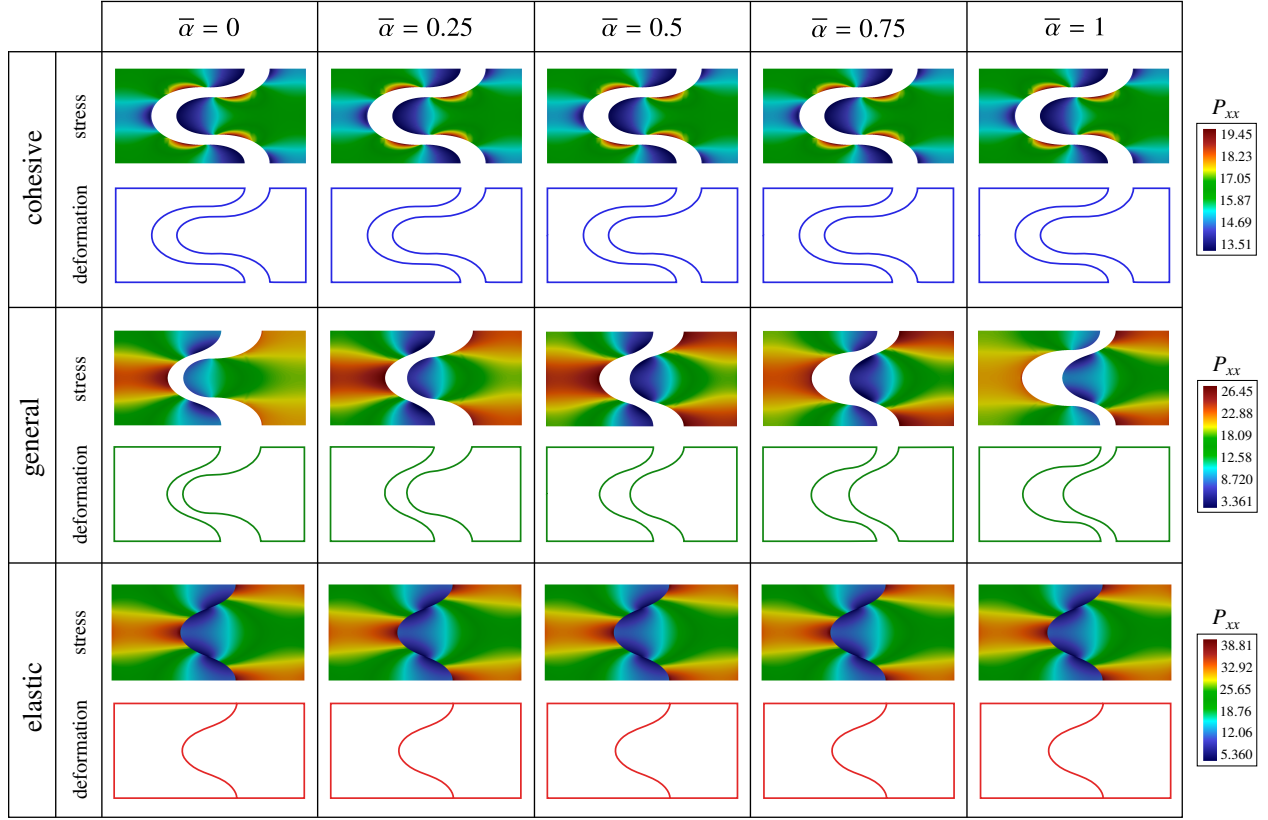}
    \caption{
    Deformations and induced stresses within the sample as it is pulled from the left and right. The results are obtained for various values of $\o{\alpha}$. The first, second and third row represent the results obtained from the cohesive, general and elastic interface models.}
    \label{fig:RES1}
\end{figure}

Next, in Fig.~\ref{fig:RES1}, we compare the induced stresses and deformations of the sample for different interface models as well as various values of $\o{\alpha}$.
The results illustrated in the first, second and third rows are obtained from the cohesive, general and elastic interface models, respectively.
As expected, the cohesive interface model allows for opening of the interface while preserving the traction continuity across the interface.
The elastic interface model, on the other hand, leads to a coherent displacement but suffers a traction jump.
The general interface model however leads to both displacement and traction jumps across the interface.
A closer look to the numerical results reveals that the results from the cohesive and elastic interfaces remain unchanged with varying the interface position while the results corresponding to the general interface model exhibit significant dependence on $\o{\alpha}$.
Such behavior is in agreement with our expectations and may be well explained as follows.
For the cohesive interface model, the only determining factor is the displacement jump across the interface.
Obviously, the interface position plays no role in evaluating the displacement jump and as a result, $\o{\alpha}$ does not contribute in governing the interface behavior in orthogonal direction.
On the other hand, the elastic interface model relies critically on the interface deformation gradient that is directly linked with $\o{\b{x}}$ and $\o{\alpha}$ as $\o{\b{F}} = \o\Grad \, \o{\b{x}} = \o\Grad \, \avg{\b{x}}_{\o{\alpha}}$.
However, due to the vanishing displacement jump across the interface, the interface placement $\o{\b{x}}$, and as a result also the interface deformation gradient remains independent of $\o{\alpha}$.
It can therefore be concluded that the influence of $\o{\alpha}$ emerges only when the interface behavior depends on the interface deformation gradient in the presence of a displacement jump.
Therefore, as confirmed by the numerical results, \emph{only the general interface shows sensitivity with respect to changing the position of the interface}.

\section{Concluding remarks}

\noindent The controversial issue of the interface position particularly relevant for the general interface model has been carefully examined and a novel view on the problem is provided.
It is commonly accepted that the interface must coincide with the mid-layer in order to satisfy the interface balance of angular momentum.
Motivated by the fact that an equivalent interface model of a graded interphase cannot coincide with the mid-layer, we have revisited this notion.
To do so, we allow the interface to assume an arbitrary position via a weighted average operator and its consequences are elaborated.
The balance laws of the general interface model based on the weighted averages are established.
It is rigorously proven that the assumption of the interface position lying on the mid-layer is not necessary to fulfill the angular momentum balance.
It is shown that the weighted average operator leads to a novel and generic form of the traction-separation law that intrinsically accounts for the position of the interface.
Moving forward, through a series of numerical examples, the general interface model is compared with its associated interphase and their relative error $\c{E}$ is calculated.
Our numerical results reveal that (i) unlike the cohesive and elastic interface models, the general interface model shows a significant sensitivity with respect to the interface position, (ii) the interface position to accurately capture a symmetrically graded interphase coincides with the mid-layer but, in general, the assumption of restricting the interface position to the mid-layer yields inaccurate results, (iii) to properly capture a graded interphase behavior, the interface position should be chosen on the stiffer part of the interphase but not on the mid-layer and (iv) for any interphase, a unique set of parameters for the general interface model exists that minimizes the error $\c{E}$.
Our next immediate plan is to extend this optimization problem to calculate $\o{\alpha}$ accounting for the points in the entire domain and not only on the interphase.
Another interesting extension is to allow for varying $\o{\alpha}$ along the interface itself.
In summary, the presented study shed lights on the potential of the general interface model and elaborates on the long standing question of the interface position.
We believe that this framework can significantly enhance our understanding of interface modeling and opens a new avenue towards designing materials with tailored structures at different scales.

\section*{Acknowledgment}
\noindent S. Saeb and P. Steinmann gratefully acknowledge the support provided by the DFG, grant number STE 544/62-2.
A. Javili gratefully acknowledge the support provided by Scientific and Technological Research Council of Turkey (T{\"U}BITAK) Career Development Program, grant number 218M700.

\appendix

\section{Proofs and intermediate steps}\label{App1}

\subsection{Proof of Eq.~(\ref{eq:4})}\label{App1-1}

\noindent The balance of moments acting on the body with respect to an arbitrary point in space yields
\begin{equation}
\begin{aligned}
&\int_{\p\c{V}\rfr} \b{x} \times \b{t}\rfr \, \d A + \int_{\c{V}\rfr} \b{x} \times \b{b}\rfr \, \d V 
\\&\quad= \int_{\p\c{V}\rfr} \b{x} \times \b{P}\cdot\b{N} \, \d A + \int_{\c{V}\rfr} \b{x} \times \b{b}\rfr \, \d V
\\&\quad\quad=\int_{\c{V}\rfr} \Div[\b{x} \times \b{P}] \, \d V + \int_{\c{V}\rfr} \b{x} \times \b{b}\rfr \, \d V
\\&\quad\quad\quad=\int_{\c{V}\rfr} \b{\varepsilon} : [\b{F} \cdot \b{P}\trns] \, \d V + \underbrace{\int_{\c{V}\rfr} \b{x} \times \Div\b{P} \, \d V + \int_{\c{V}\rfr} \b{x} \times \b{b}\rfr \, \d V}_{\text{=$\b{0}$ due to \eqref{eq:2}}} = \b{0}  \quad \forall \c{V}\rfr \\
&\Rightarrow \quad 	\b{\varepsilon} : [\b{F} \cdot \b{P}\trns] = \b{0} \,.
\end{aligned}
\end{equation}

\subsection{Proof of Eq.~(\ref{eq:6})}\label{App1-2}

\noindent In the material configuration and in the limit of localizing the cutout volume to the interface, the boundaries $\p\c{V}^{+}\rfr$ and $\p\c{V}^{-}\rfr$ coincide geometrically with the interface and therefore, ${\p\c{V}^{-}\rfr}={\p\c{V}^{+}\rfr} = {\c{I}\rfr}$.
However, in this limit, the normal vector on $\p\c{V}\rfr^{+}$ coincides with the interface normal vector $\o{\b{N}}$ while the the normal vector on $\p\c{V}\rfr^{-}$ points in the opposite direction.
Equation~\eqref{eq:4.5} is therefore rewritten as
\begin{equation}
\begin{aligned}
&\int_{\c{I}\rfr} \b{P}^{+} \cdot \o{\b{N}} \, \d A - \int_{\c{I}\rfr} \b{P}^{-} \cdot \o{\b{N}} \, \d A +  \int_{\p\c{I}\rfr} \o{\b{P}} \cdot \t{\b{N}} \, \d L + \int_{\c{I}\rfr} \o{\b{b}}\rfr \, \d A
\\&\quad= \int_{\c{I}\rfr} \llbracket \b{P} \rrbracket \cdot \o{\b{N}} \, \d A +  \int_{\p\c{I}\rfr} \o{\b{P}} \cdot \t{\b{N}} \, \d L + \int_{\c{I}\rfr} \o{\b{b}}\rfr \, \d A
\\&\quad\quad=\int_{\c{I}\rfr} \llbracket \b{P} \rrbracket \cdot \o{\b{N}} \, \d A +  \int_{\c{I}\rfr} \o{\Div}\,\o{\b{P}} \, \d A + \int_{\c{I}\rfr} \o{C}\,\underbrace{\o{\b{P}}\cdot\o{\b{N}}}_{=\b{0}} \, \d A  + \int_{\c{I}\rfr} \o{\b{b}}\rfr \, \d A = \b{0}\,,{eq:6}
\end{aligned}
\end{equation}
where the third integral vanishes as a result of the superficiality of the interface Piola stress.
Due to the arbitrariness of the cutout volume, the strong form of the balance of forces on the interface reads
\begin{equation}
\o{\Div} \, \o{\b{P}} + \llbracket \b{P} \rrbracket \cdot \o{\b{N}} + \o{\b{b}}\rfr = \b{0} \, .
\end{equation}

\subsection{Proof of Eq.~(\ref{eq:8})}\label{App1-3}

\noindent The balance of moments acting on the interface with respect to an arbitrary point in space reads
\begin{equation}
\begin{aligned}
&\int_{\p\c{V}\rfr^{+}} \b{x} \times \b{t}\rfr \, \d A + \int_{\p\c{V}\rfr^{-}} \b{x} \times \b{t}\rfr \, \d A +  \int_{\p\c{I}\rfr} \o{\b{x}} \times \o{\b{P}} \cdot \t{\b{N}} \, \d L + \int_{\c{I}\rfr} \o{\b{x}} \times \o{\b{b}}\rfr \, \d A
\\&\quad=\int_{\c{I}\rfr} \b{x}^{+} \times \b{P}^{+} \cdot \o{\b{N}} \, \d A - \int_{\c{I}\rfr} \b{x}^{-} \times \b{P}^{-} \cdot \o{\b{N}} \, \d A +  \int_{\p\c{I}\rfr} \o{\b{x}} \times \o{\b{P}} \cdot \t{\b{N}} \, \d L + \int_{\c{I}\rfr} \o{\b{x}} \times \o{\b{b}}\rfr \, \d A
\\&\quad\quad=\int_{\c{I}\rfr} \llbracket \b{x} \times \b{P} \rrbracket \cdot \o{\b{N}} \, \d A +  \int_{\c{I}\rfr} \o{\Div}\, [\o{\b{x}} \times \o{\b{P}}] \, \d A + \int_{\c{I}\rfr} \o{C}\, \o{\b{x}} \times\,\underbrace{\o{\b{P}}\cdot\o{\b{N}}}_{=\b{0}} \, \d A + \int_{\c{I}\rfr} \o{\b{x}} \times \o{\b{b}}\rfr \, \d A
\\&\quad\quad\quad=\int_{\c{I}\rfr} \llbracket \b{x} \times \b{P} \rrbracket \cdot \o{\b{N}} \, \d A +  \int_{\c{I}\rfr} \b{\varepsilon} : [\o{\b{F}} \cdot \o{\b{P}}\trns] \, \d A +  \int_{\c{I}\rfr} \o{\b{x}} \times \o{\Div}\, \o{\b{P}} \, \d A + \int_{\c{I}\rfr} \o{\b{x}} \times \o{\b{b}}\rfr \, \d A
\\&\quad\quad\quad\quad=\int_{\c{I}\rfr} \llbracket \b{x} \times \b{P} \rrbracket \cdot \o{\b{N}} \, \d A +  \int_{\c{I}\rfr} \b{\varepsilon} : [\o{\b{F}} \cdot \o{\b{P}}\trns] \, \d A - \int_{\c{I}\rfr} \o{\b{x}} \times \llbracket \b{P} \rrbracket \cdot \o{\b{N}} \, \d A = \b{0}   \quad \forall \c{V}\rfr \, ,
\end{aligned}
\end{equation}
that holds for any cutout volume.
Consequently, the interface balance of moments after localization is obtained as
\begin{equation}
\llbracket\b{x} \times \b{P} \rrbracket \cdot \o{\b{N}} + \b{\varepsilon} : [\o{\b{F}} \cdot \o{\b{P}}\trns] - \o{\b{x}} \times \llbracket \b{P} \rrbracket \cdot \o{\b{N}} = \b{0} \, .
\end{equation}

\subsection{Weak form of the governing equations}\label{weak-mech}

\noindent The primary step towards the finite element implementation of the theory is to derive the weak form of the governing equations.
The balances of moments in the bulk and on the interface do not contribute to the weak formulation of the governing equations as they are fulfilled through choosing suitable material models.
Therefore, only the weak form of the balances of forces is established.
To do so, the left hand side of Eqns.~$\eqref{eq:2}$ and $\eqref{eq:6}$ are tested with vector-valued test functions, $\delta\b{\varphi}\in \c H\rfr^{1}(\c{B}\rfr)$ and $\delta{\o{\b{\varphi}}}\in \c H\rfr^{1}(\c{I}\rfr)$, respectively and are integrated over the bulk and interface domain in the material configuration.
Through combining the resulting equations and employing the extended divergence theorem along with the definition $\delta\o{\b{\varphi}} := \avg{\delta{\b{\varphi}}}_{\o{\alpha}}$, we obtain
\begin{equation}\label{weak-form}
\begin{aligned}
&\int _{\c V\rfr} \delta\b\varphi \cdot \Div \b P \, \d V +
\int _{\c V\rfr} \delta\b\varphi \cdot {\b b}\rfr \, \d V +
\int_{\c{I}\rfr} {\delta \o{\b\varphi}} \cdot\o\Div\,\o{\b{P}} \, \d A +
\int_{\c{I}\rfr} \delta\o{\b{\varphi}} \cdot \llbracket \b{P} \rrbracket \cdot \o{\b{N}} \, \d A +
\int_{\c{I}\rfr} \delta\o{\b{\varphi}} \cdot \o{\b{b}}\rfr \, \d A
\\&\quad=\int _{\c V\rfr} \Div [\delta\b\varphi \cdot \b P] \, \d A -
\int _{\c V\rfr} \b P : \Grad \delta\b\varphi \, \d V+
\int _{\c V\rfr} \delta\b\varphi \cdot {\b b}\rfr \, \d V
\\&\quad\quad+\int_{\p\c{I}\rfr} {\delta \o{\b\varphi}} \cdot \o{\b{t}}\rfr \, \d L -
\int_{\c{I}\rfr} \o{\b{P}}:\o{\Grad}\,{\delta \o{\b\varphi}} \, \d A +
\int_{\c{I}\rfr} \delta\o{\b{\varphi}} \cdot \llbracket \b{P} \rrbracket \cdot \o{\b{N}} \, \d A +
\int_{\c{I}\rfr} \delta\o{\b{\varphi}} \cdot \o{\b{b}}\rfr \, \d A
\\&\quad\quad\quad=\int _{\p\c B\rfr} \delta\b\varphi \cdot {\b t}\rfr \, \d A -
\int _{\c I\rfr} \llbracket \delta\b\varphi \cdot \b t\rfr \rrbracket \, \d A -
\int _{\c B\rfr} \b P : \Grad \delta\b\varphi \, \d V +
\int _{\c B\rfr} \delta\b\varphi \cdot {\b b}\rfr \, \d V \\
&\quad\quad\quad\quad+\int_{\p\c{I}\rfr} \lravg{\delta\b{\varphi}}_{\o{\alpha}} \cdot \o{\b{t}}\rfr \, \d L-
\int_{\c{I}\rfr} \o{\b{P}}:\o{\Grad}\,\avg{\delta{\b{\varphi}}}_{\o{\alpha}} \, \d A +
\int_{\c{I}\rfr} \lravg{\delta\b{\varphi}}_{\o{\alpha}} \cdot \llbracket \b{t}\rfr \rrbracket \, \d A +
\int_{\c{I}\rfr} \lravg{\delta\b{\varphi}}_{\o{\alpha}} \cdot  \o{\b{b}}\rfr \, \d A = \b{0} \, .
\end{aligned}
\end{equation}
Next, we employ the identity \eqref{magic-identity-ex} to rewrite the second integral in terms of the jumps, weighted average and complementary weighted average of the integrands.
That is
\begin{equation}\label{weak-form-1}
\begin{aligned}
&\int _{\p\c B\rfr} \delta\b\varphi \cdot {\b t}\rfr \, \d A -
\int _{\c I\rfr} \avg{\delta\b{\varphi}}_{\o{\alpha}} \cdot \llbracket \b t \rrbracket \, \d A -
\int _{\c I\rfr} \llbracket {\delta\b{\varphi}} \rrbracket \cdot \avg{ \b t }_{1-\o{\alpha}} \, \d A -
\int _{\c B\rfr} \b P : \Grad \delta\b\varphi \, \d V +
\int _{\c B\rfr} \delta\b\varphi \cdot {\b b}\rfr \, \d V \\
&\quad+\int_{\p\c{I}\rfr} \lravg{\delta\b{\varphi}}_{\o{\alpha}} \cdot \o{\b{t}}\rfr \, \d L-
\int_{\c{I}\rfr} \o{\b{P}}:\o{\Grad}\,\avg{\delta\b{\varphi}}_{\o{\alpha}} \, \d A +
\int_{\c{I}\rfr} \lravg{\b{\varphi}}_{\o{\alpha}} \cdot \llbracket \b{t} \rrbracket \, \d A +
\int_{\c{I}\rfr} \lravg{\delta\b{\varphi}}_{\o{\alpha}} \cdot \o{\b{b}}\rfr \, \d A
\\&\qquad=\underbrace{ \int _{\p\c B\rfr} \delta\b\varphi \cdot {\b t}\rfr \, \d A -
\int _{\c B\rfr} \b P : \Grad \delta\b\varphi \, \d V +
\int _{\c B\rfr} \delta\b\varphi \cdot {\b b}\rfr \, \d V}_{\text{bulk}}
\\&\qquad\quad+\underbrace{
\int_{\p\c{I}\rfr} \lravg{\delta\b{\varphi}}_{\o{\alpha}} \cdot \o{\b{t}} \, \d L-
\int_{\c{I}\rfr} \o{\b{P}}:\o{\Grad}\,\lravg{\delta\b{\varphi}}_{\o{\alpha}} \, \d A +
\int_{\c{I}\rfr} \lravg{\delta\b{\varphi}}_{\o{\alpha}} \cdot \o{\b{b}}\rfr \, \d A -
\int _{\c I\rfr} \llbracket {\delta\b{\varphi}} \rrbracket \cdot \avg{ \b t }_{1-\o{\alpha}} \, \d A
\overset{.}{=} \b{0} \, }_{\text{interface}}.
\end{aligned}
\end{equation}
Equation~\eqref{weak-form-1} is a generic weak form of the balance of forces and allows for any arbitrary position of the interface between its two sides.
Obviously, for $\o{\alpha}=\tfrac{1}{2}$, the standard form of the weak form for balance of forces on the interface \citep{Javili2017c} is recovered.


\bibliography{library}

\end{document}